\documentclass[prd,twocolumn,showpacs,floatfix,amsmath,nofootinbib,amssymb,floatfix]{revtex4}
\usepackage{graphicx,color,dcolumn,booktabs,bm,multirow}
\usepackage{longtable,lscape}
\usepackage{txfonts}
\usepackage{CJK}
\usepackage{overpic}
\usepackage{amssymb}
\usepackage{feynmf}   
\usepackage{slashed}  
\usepackage{cases}
\usepackage{multirow}
\usepackage{epstopdf}
\usepackage{amsfonts}
\usepackage{subfigure}
\usepackage{graphics}
\usepackage{chngpage}
\usepackage{array}
\usepackage[colorlinks,
            citecolor=green,
            anchorcolor=red,
            menucolor=red,
            linkcolor=red,
            filecolor=red,
            runcolor=red,
            urlcolor=blue,
            frenchlinks=red]{hyperref}

\begin{document}
\begin{CJK*}{GB}{gbsn}
\title{Possible hadronic molecular states composed of $S$-wave heavy-light mesons}
\author{Ming-Zhu Liu(ÁõÃ÷Öé)$^1$}
\author{Duo-Jie Jia(¼Ö¶à½Ü)$^1$}\email{ jiadj@nwnu.edu.cn}
\author{Dian-Yong Chen(³ÂµîÓÂ)$^{2}$ \footnote{Corresponding author}} \email{chendy@seu.edu.cn}
\affiliation{$^1$ Institute of Theoretical Physics, College of Physics and Electronic Engineering, Northwest Normal University, Lanzhou 730070, People's Republic of China\\
$^2$Department of Physics, Southeast University, Nanjing 210094, People's Republic of China}
\date{\today}
\begin{abstract}
We perform a systematical  study of
possible molecular states composed of the $S$ wave heavy light
mesons, where the $S-D$ mixing and {$\eta-\eta^{\prime}$
mixing} are explicitly included. Our calculation
indicates that the observed $X(3872)$ could be a loosely shallow
molecular state composed of $D\bar{D}^\ast +h.c$, while neither
$Z_c(3900)/Z_c(4020)$ nor $Z_b(10610)/Z_b(10650)$ is supported to
be a molecule. Some observed possible molecular states are
predicted, which could be searched for by further experimental
measurements.
\end{abstract}
\maketitle

\section{Introduction}
In the past decade, a number of new hadron states, named $XYZ$ particles, have been observed experimentally \cite{Chen:2016qju}. Among these newly observed states, some of them are close to the thresholds of a pair of hadrons, which indicates these kind of new states could be good candidates for hadronic molecular states. A typical example of the new hadron states, $X(3872)$, was first observed by the Belle Collaboration in the $\pi^+ \pi^- J/\psi$ invariant mass of the $B^\pm \to KX(3972) \to K(\pi^+ \pi^- J/\psi)$ process in 2003 \cite{Choi:2003ue}. Later, this state was successfully confirmed by Belle itself \cite{Abe:2005ix,Gokhroo:2006bt,:2008te,:2008su} and by the  Babar \cite{Aubert:2004ns,Aubert:2004fc,Aubert:2005eg,Aubert:2005zh,Aubert:2005vi,Aubert:2006aj,Aubert:2007rva}, CDF \cite{Acosta:2003zx,Abulencia:2005zc,Abulencia:2006ma,Aaltonen:2009vj}, D0 \cite{Abazov:2004kp}, LHCb \cite{Aaij:2011sn, Aaij:2013zoa, Aaij:2014ala, Aaij:2015eva} and BESIII \cite{Ablikim:2013dyn} Collaborations in the $\pi^+ \pi^- \pi^0 J/\psi $, $D^0 \bar{D}^0 \pi$, $D^{\ast0} \bar{D}^0$, $\gamma J/\psi$ and $\gamma \psi(2S)$ processes. The $J^{PC}$ quantum numbers of the $X(3872)$ have been confirmed as $1^{++}$ and the PDG average of the mass and width are $3871.69 \pm 0.17$ and $<1.2$ MeV, respectively. The observed mass of the $X(3872)$ is just sandwiched by the thresholds of the $D^{\ast 0} \bar{D}^0$ and $D^{\ast +} {D}^-$. The absence of the charge partner of the $X(3872)$ indicates that this state is an isospin singlet \cite{Aubert:2004zr}.

A very similar charmonium-like state to the $X(3872)$, the $Z_c (3900)$, was first reported by the BESIII and Belle Collaborations in the $\pi^{\pm} J/\psi $ invariant mass spectrum of the $e^+ e^- \to \pi^+ \pi^- J/\psi $ at a center-of-mass energy of 4.260 GeV \cite{Ablikim:2013mio, Liu:2013dau}. Later, this state was confirmed at the same process but at $\sqrt{s}=4.17$ GeV by the CLEO Collaboration \cite{Xiao:2013iha}. The open charm decay channel $Z_c(3900) \to D^\ast \bar{D}$ was reported by the BESIII Collaboration  in 2014 \cite{Ablikim:2013xfr}. Recently, the neutral partner of the $Z_c(3900)^\pm$ has been observed in the $\pi^0 J/\psi$ and $(D\bar{D}^\ast)^0$ invariant mass spectra by the CLEO \cite{Xiao:2013iha} and BESIII Collaborations \cite{Ablikim:2015tbp, Ablikim:2015gda}. As an isospin triplet, the mass of the $Z_c(3900)$ is very close to the  threshold of the $D\bar{D}^\ast$. As a partner of the $Z_c(3900)$, $Z_c(4020)$ is very close to the threshold of the $D^\ast \bar{D}^\ast$, which was first observed in the $\pi^\pm h_c$ invariant mass spectrum by the BESIII Collaboration \cite{Ablikim:2013wzq, Ablikim:2014dxl} then confirmed in the $D^\ast \bar{D}^\ast$ invariant mass spectrum \cite{Ablikim:2013emm, Ablikim:2015vvn}.

All three states, $X(3872)$, $Z_c(3900)$ and $Z_c(4020)$,
are close to the thresholds of a pair of charmed mesons and
were first observed in hidden charm processes. States
near the $D_s^{(\ast)+} D_s^{(\ast)-}$ { thresholds} should
be more easily discovered in the hidden charm process with a light
meson containing $s\bar{s}$ quark components, due to the simple
quark rearrangement. A series of charmonium-like
states have recently been observed in the $J/\psi \phi$ invariant mass
spectrum of the $B\to K J/\psi \phi$ process, the $Y(4140)$,
$Y(4274)$, $X(4320)$ and $X(4350)$ \cite{Aaltonen:2009tz,
Aaltonen:2011at, Chatrchyan:2013dma, Abazov:2013xda, Aaij:2016iza,
Aaij:2016nsc}. Among these states, the $Y(4140)$ is about 80 MeV
below the threshold of the $D_s^{\ast+ } D_s^{\ast -}$. Here, one
should notice that the thresholds of the $D_s^{\ast +}D_s^{-}$ and
$D_s^{+}D_s^{-}$ are below that of the $J/\psi \phi$, thus, one
cannot observe the states near the $D_s^{\ast +}D_s^{-}$ and
$D_s^{+}D_s^{-}$ thresholds in the $J/\psi \phi$ invariant mass
spectrum.

In the bottom sector, $Z_b(10610)$ and $Z_b(10650)$ were first reported in the $\Upsilon(nS) \pi^\pm,\ \{n=1,2,3\}$ and $h_b(mP) \pi^\pm\ \{m=1,2\}$ invariant mass spectra of the $e^+ e^- \to \Upsilon(nS) \pi^+ \pi^-$ and $e^+ e^- \to h_b(mP) \pi^+ \pi^-$ process at a center-of-mass energy of 10.860 GeV by the Belle Collaboration in 2011 \cite{Belle:2011aa, Garmash:2014dhx}. The open bottom channels of the $Z_b(10610)$ and $Z_b(10650)$ were observed by Belle in 2012 \cite{Adachi:2012cx, Garmash:2015rfd}. The neutral partners of the $Z_b(10610)$ and $Z_b(10650)$ were observed in the hidden bottom channel in 2013 \cite{Krokovny:2013mgx}. More recently, the signals of these two bottom-like states have also been discovered in the $e^+ e^- \to h_b(mP) \pi^+ \pi^-$ at a center-of-mass energy of 11.020 GeV \cite{Abdesselam:2015zza}.

The observed masses of the $Z_b(10610)$ and $Z_b(10650)$ are very close to the thresholds of the $B^\ast \bar{B}$ and $B^\ast \bar{B}^\ast$, respectively, and could be considered as corresponding to the charmonium-like states $Z_c(3900)$ and $Z_c(4020)$ in the bottom sector. As the bottom counterpart of the $X(3872)$,  it was proposed to search for the $X_b$ in the $\Upsilon \pi^+ \pi^-$ process \cite{Hou:2006it}, hidden bottom decay channels \cite{Li:2015uwa} and the radiative decay of the $\Upsilon(5S)$ and $\Upsilon(6S)$ \cite{Wu:2016dws}. The Belle Collaboration searched for the signal of the $X_b$ in the $\omega \Upsilon(1S)$ channel and found no evidence of the $X_b$ state \cite{He:2014sqj}.

To date, four charmonium-like states, $X(3872)$, $Z_c(3900)$, $Z_c(4020)$ and $Y(4140)$, have been observed experimentally, and are near the thresholds of a pair of $S$ wave charmed or charmed-strange mesons. In the bottom sector, two bottomonium-like states, $Z_b(10610)$ and $Z_b(10650)$, have been discovered which could be the bottom counterparts of the $Z_c(3900)$ and $Z_c(4020)$. In Table \ref{Tab:threshold}, we summarize the thresholds of pairs of $S$-wave charmed/charmed-strange and bottom/bottom-strange mesons, and the corresponding near-threshold states are also presented.

\begin{table}
\caption{The thresholds of pairs of $S$-wave heavy-light mesons
and the corresponding near-threshold charmonium-like
states.\label{Tab:threshold}}
\begin{tabular}{c|cccc}
\toprule[1pt]\toprule[1pt]
\multicolumn{2}{c}{} & Threshold & \multicolumn{2}{c}{Possible State} \\
\cline{4-5}
\multicolumn{2}{c}{} & (MeV)       & Isospin Singlet & Isospin triplet \\
\midrule[1pt] \multirow{9}{*}{\rotatebox{90}{charm Sector}}
             & $D\bar{D}$                  &3734        &$\cdots$        &$\cdots$\\
             & $D^\ast \bar{D}$            &3876        &$X(3872)$       &$Z_c(3900)$ \\
             & $D^\ast \bar{D}^\ast$       &4017        &$\cdots$        &$Z_c(4020)$\\
             & $D_s^+D_s^-$                &3936        &$\cdots$        &$\cdots$\\
             & $D_s^{\ast+} {D}_s^-$       &4080        &$\cdots$        &$\cdots$ \\
             & $D_s^{\ast+} {D}_s^{\ast-}$ &4224        &$\cdots$        &$Y(4140)$\\
             & $D D_s$                     &3835        &$\cdots$        &$\cdots$\\
             & $D D_s^{\ast}$              &3979        &$\cdots$        &$\cdots$\\
             & $D^\ast D_s$                &3977        &$\cdots$        &$\cdots$\\
             & $D^\ast D_s^{\ast}$         &4121        &$\cdots$        &$\cdots$\\
\midrule[1pt] \hspace{3mm}\multirow{9}{*}{\rotatebox{90}{bottom
Sector}}\hspace{3mm}
             & $B\bar{B}$                      &10559            &$\cdots$        &$\cdots$\\
             & $B^\ast \bar{B}$                &10605            &$\cdots$        &$Z_b(10610)$ \\
             & $B^\ast \bar{B}^\ast$           &10651            &$\cdots$        &$Z_b(10650)$\\
             & $B_s^0\bar{B}_s^0$              &10734            &$\cdots$        &$\cdots$\\
             & $B_s^{\ast0} \bar{B}_s^0$       &10782            &$\cdots$        &$\cdots$\\
             & $B_s^{\ast0} \bar{B}_s^{\ast0}$ &10830            &$\cdots$        &$\cdots$\\
             & $B \bar{B}_s^0$                 &10646            &$\cdots$        &$\cdots$\\
             & $B \bar{B}_s^{0\ast}$           &10694            &$\cdots$        &$\cdots$\\
             & $B^\ast \bar{B}_s^0$            &10694            &$\cdots$        &$\cdots$\\
             & $B^\ast \bar{B}_s^{\ast0}$      &10740            &$\cdots$        &$\cdots$\\

\bottomrule[1pt]\bottomrule[1pt]
\end{tabular}
\end{table}

The experimental observations have stimulated theorists to great interest in the intrinsic nature of these near-threshold states. Different interpretations of these observed states have been proposed, such as conventional charmonium to the $X(3872)$ and $Y(4140)$
\cite{Kalashnikova:2005ui, Zhang:2009bv, Kalashnikova:2009gt, Li:2009zu,Ferretti:2013faa, Eichten:2004uh, Pennington:2007xr, Chen:2016iua},
tetraquark states \cite{Maiani:2004vq, Chiu:2006us, Ebert:2005nc,Ali:2014dva,Ali:2013xba,Chen:2010ze,Chen:2015ata,
Chen:2013omd,Wang:2013llv,Qiao:2013dda,Wang:2015nwa} and special production mechanisms \cite{Bugg:2011jr, Bugg:2011ub, Chen:2011pv, Chen:2013coa, Chen:2011xk, Chen:2011zv, Chen:2013wca}. Since all the above-mentioned states are near-threshold states, the hadronic molecular interpretations of these states are particularly attractive. In the following, we present a short review of the molecular interpretations of the observed near-threshold states listed in Table \ref{Tab:threshold}.

{\it Molecular interpretation of $X(3872)$:--} The first observed charmonium-like state, $X(3872)$, is very close to the threshold of the $D^\ast \bar{D}$, so it is natural to consider the $X(3872)$ as a shallow bound state of the $D^\ast \bar{D }+ h.c$. In Ref. \cite{Swanson:2003tb}, the author proposed a microscopic model, where both the quark exchange and pion exchange induced effective potential were included and  the $X(3872)$ was interpreted as a $D^{\ast 0} \bar{D}^0 +h.c$ molecular state. The calculation at  the quark level suggested that molecular states $D^{0}\bar{D}^{\ast0}$,
$D^{+}\bar{D}^{\ast-}$ and $D^{-}\bar{D}^{\ast+}$ could be mixed to form components of $I=0$ and $I=1$ states, and the $I=0$ state could correspond to the observed $X(3872)$. The one-boson-exchange potential model calculations indicated that the $X(3872)$ could be a $D^\ast \bar{D} +h.c $ molecular state \cite{Liu:2008fh, Thomas:2008ja,Liu:2008tn}. The estimation by the effective Lagrangian \cite{AlFiky:2005jd}, coupled channel \cite{Lee:2009hy} and QCD sum rule \cite{Chen:2013pya} also supported the $X(3872)$ as a shallow $\bar{D}^{\ast}\bar{D}$ bound state.

In the molecular framework the decay behaviors of the $X(3872)$ have been extensively discussed. In Refs. \cite{Dong:2009yp, Dong:2014zka, Harada:2010bs}, the strong and radiative decays of the $X(3872)$ were discussed in the $D^{\ast0} \bar{D}^{0}+h.c$ molecular scenario with the compositeness condition of the composite particle. The estimate in an effective Lagrangian approach in the molecular scenario were consistent with the corresponding experimental measurement \cite{Braaten:2010mg}, which indicated that the $X(3872)$ could be a loosely bound state of the $D\bar{D}^\ast + h.c$.

{\it Molecular interpretation of $Z_c(3900)$ and $Z_c(4020)$:--} The observed masses of the $Z_c(3900)$ and $Z_c(4020)$ are very close to the thresholds of $D^\ast \bar{D}$ and $D^\ast \bar{D}^\ast$, respectively, which indicates that the they could be a $D^\ast \bar{D}$ and $D^\ast \bar{D}^\ast$ hadronic molecular state with $I=1$. The authors of Refs. \cite{Sun:2011uh, Sun:2012zzd, Aceti:2014uea} used the potential model to find  bound state solutions for the $D^\ast \bar{D} +h.c$ and $D^\ast \bar{D}^\ast $systems, which corresponded well to the observed $Z_c(3900)$ and $Z_c(4020)$. The QCD sum rule calculations in Refs. \cite{Wang:2013daa, Chen:2015ata} also supported that the $Z_c(3900)$ and $Z_c(4020)$ could be deuteron-like hadronic molecular states.

The decays of the $Z_c(3900)$ and $Z_c(4020)$ were estimated via the meson loops \cite{Li:2013xia, Wu:2016ypc}. The product and decay behaviors of $Z_c(3900)$ and $Z_c(4020)$ have been studied in $D^\ast \bar{D}+h.c$ and $D^\ast \bar{D}^\ast $ hadronic molecular scenarios with the Weinberg compositeness condition in Refs. \cite{Dong:2013iqa, Dong:2013kta}, and the theoretical estimations were consistent with the corresponding experimental measurements. Besides the observed channels, some other decay modes of $Z_c(3900)$ and $Z_c(4020)$ have been studied in the molecular scenario, such as the $\rho \eta_c$, $J/\psi \pi \gamma$, $\gamma \eta_c$ and $\gamma \chi_{cJ}$ \cite{Li:2014pfa, Gutsche:2014zda, Esposito:2014hsa, Ke:2013gia, Chen:2015igx}. All these theoretical studies supported that the $Z_c(3900)$ and $Z_c(4020)$ could be assigned as $D^\ast \bar{D} +h.c$ and $D^\ast \bar{D}^\ast$ hadronic molecular states, respectively.

{\it Molecular interpretation of $Y(4140)$:--} The observed mass of the $Y(4140)$ is about 80 MeV below the thresholds of the $D_s^{\ast +} D_s^{\ast -}$, and the charmed-strange meson
pair could easily couple to the $J/\psi \phi$ final states, so it is natural to interpret the $Y(4140)$ as a $S-$ wave $D_s^{\ast +} D_s^{\ast -}$ molecule. The potential calculations in Refs \cite{Mahajan:2009pj,Ding:2009vd, Liu:2009ei,  Molina:2009ct, Liu:2009pu, Zhang:2009vs} indicated that the $Y(4140)$ could be a $D_s^{\ast+} {D}_s^{\ast-}$ molecular state with $J^{PC}=0^{++}$.  The QCD sum rule calculations also supported the $Y(4140)$ to be a $D_s^{\ast +} D_s^{\ast-}$ molecular state \cite{Albuquerque:2009ak, Zhang:2009st, Wang:2014gwa}. In Ref. \cite{Chen:2016ugz}, $Y(4140)$ was assigned as a mixing $D_s^{\ast +} D_s^{\ast -}$ molecular state with $D^{\ast} \bar{D}^{\ast}$ component.

The lineshape of the radiative open-charm decay of the $Y(4140)$ is estimated in Ref. \cite{Liu:2009pu}, where the $Y(4140)$ was considered as the strange counterpart of the $Y(3930)$. The hidden charm decays of the $Y(4140)$ were studied in the hadronic molecular state \cite{Branz:2009yt} with $J^{PC}=0^{++}$ and $2^{++}$.

{\it Molecular interpretation of $Z_b(10610)$ and $Z_b(10650)$:--}
The experimentally measured masses of the $Z_b(10610)$ and $Z_b(10650)$ are very close to the thresholds of the $B^\ast \bar{B}$ and $B^\ast \bar{B}^\ast$. In Refs. \cite{Liu:2008fh,Sun:2011uh}, the OBE potential model indicated that the
$Z_{b}(10610)$ and $Z_{b}(10650)$ could be molecular states composed of $B\bar{B}^{\ast}$ and $B^{\ast}\bar{B}^{\ast}$,
respectively. The observed $Z_b(10610)$ and $Z_b(10650)$ were explained as molecular states in the chiral quark model \cite{Li:2012wf,Yang:2011rp}. Using QCD sum rules, the masses of the $Z_b(10610)$ and $Z_b(10650)$ could be reproduced in a molecular picture \cite{Zhang:2011jja, Wang:2013daa, Wang:2014gwa}.

In Ref~\cite{Dong:2012hc}, the transitions to $\Upsilon(nS)\pi\ (n=1,2,3)$ and $h_b(mP) \pi \ (m=1,2)$ were analysed in the molecular picture with compositeness condition. The observed processes of the $Z_b(10610)$ and $Z_b(10650)$ investigated in the effective Lagrangian approach also supported the molecular scenarios \cite{Li:2012as, Cleven:2013sq, Li:2012uc,Mehen:2011yh}. The decays of the $Z_b(10610)$ and $Z_b(10650)$ have been evaluated via the intermediate meson loops model, where more decay channels were predicted \cite{Li:2014pfa}.

 In Table \ref{Tab:threshold}, there exist 10 thresholds of pairs of charmed or bottom mesons. As we discussed above, some near-threshold charmonium-like or bottomonium-like states have been observed experimentally, and have been intensively considered as $S-$ wave hadornic molecular states. Theoretically, it is very interesting and urgent to systematically consider the possibility of hadronic molecular states composed of other combinations of charmed or bottom meson pairs \cite{Liu:2008tn}. Moreover, investigations of the deuteron indicated that the $D$-wave component of the wave function is crucial in understanding its static properties \cite{Carlson:1997qn, Arnold:1980zj}. Thus, in the present work, we further include the $S-D$ mixing in the wave functions of the hadronic molecule composed of a heavy-light meson pairs. By this systematic study, we can identify whether the observed near threshold states, i.e.,  $X(3872)$, $Z_c(3900)$, $Z_c(4020)$, $Y(4140)$, $Z_b(10610)$ and $Z_b(10650)$, could be hadronic molecular states and in addition, we can predict more near-threshold molecular states, which could be accessed by further experimental measurements.

This work is organized as follows. After this Introduction, we present the wave functions of the possible molecular state and the effective potentials of the heavy-light meson pair in Section \ref{Sec:2}. The numerical results and discussion are given in Section \ref{Sec:3} and Section \ref{Sec:4} is devoted to a summary.

\section{Wave functions and effective potentials  \label{Sec:2}}

In the heavy quark effective theory, the two $S$-wave heavy-light mesons degenerate into a $H=\{\mathcal{P}, \mathcal{P}^\ast\}$ doublet, in which the $\mathcal{P}$ and $\mathcal{P}^\ast$ indicate $D_{(s)}$ and $D_{(s)}^\ast$ in the charm sector and $B_{(s)}$ and $B_{(s)}^{\ast}$ in the bottom sector. The molecular state composed of $H\bar{H}$ can be decomposed into three types, which are $\mathcal{P}-\mathcal{P}$, $\mathcal{P}-\mathcal{P}^\ast$ and $\mathcal{P}^\ast-\mathcal{P}^\ast$, respectively. In the following, we construct the wave functions and calculate the potentials of these three types.

\subsection{Wave function of the molecular states}

For a molecular state composed of two mesons, the total wave function is
\begin{eqnarray}
|\Psi\rangle=\Big|\frac{\phi(r)}{r}\Big\rangle\otimes|{}^{2S+1}L_{J}\rangle\otimes|I,I_{3}\rangle
\end{eqnarray}
where the $|\frac{\phi(r)}{r}\rangle$, $|L_{J}\rangle$and $|I,I_{3}\rangle$ denote the radial, spin-orbital and flavor
functions, respectively. As for the radial and the spin-orbital wave function, there exists $S-D$ mixing in the $\mathcal{P}-\mathcal{P}^{\ast}$ and $\mathcal{P}^{\ast}-\mathcal{P}^{\ast}$ types of hadronic molecular states, which will be considered explicitly in the present work. For the $S-$ wave dominant $\mathcal{P}-\mathcal{P}^{\ast}$ type molecule, both the spin and total angular momentum are one, while the orbital momentum could be zero and two when considering the $S-D$ mixing. The corresponding spin-orbital wave functions are
\begin{eqnarray}
J=1 : |^3S_1\rangle, |^3D_1\rangle.
\end{eqnarray}
The general decomposition of the spin-orbital wave function $|^{2S+1}L_J\rangle$ for $\mathcal{P}-\mathcal{P}^{\ast}$ system is
\begin{eqnarray}
|{}^{2S+1}L_{J}\rangle=\sum_{m_s,m_{L}}C_{S m_s,Lm_{L}}^{JM}\epsilon_{n}^{m}Y_{Lm_{L}},
\end{eqnarray}
where $C_{S m_{s},Lm_{L}}^{JM}$ are Clebsch-Gordan coefficients, $Y_{Lm_{L}}$ is the spherical harmonics functions and $\epsilon_{n}^{m}$ is the polarization vector for the vector meson.

As for the $\mathcal{P}^{\ast}-\mathcal{P}^{\ast}$ type molecular states, the considered total angular momentum could be 0, 1 or 2, and the corresponding spin-orbital wave function could be
\begin{eqnarray}
J&=&0:|{}^{1}\mathbb{S}_{0}\rangle,|{}^{5}\mathbb{D}_{0}\rangle,\nonumber\\
J&=&1:|{}^{3}\mathbb{S}_{1}\rangle,|{}^{3}\mathbb{D}_{1}\rangle,|{}^{5}\mathbb{D}_{1}\rangle,\nonumber\\
J&=&2:|{}^{5}\mathbb{S}_{2}\rangle,|{}^{1}\mathbb{D}_{2}\rangle,|{}^{3}\mathbb{D}_{2}\rangle,|{}^{5}\mathbb{D}_{2}\rangle,
\end{eqnarray}
respectively. The decomposed form of the above wave function is
\begin{eqnarray}
|{}^{2S+1}L_{J}\rangle=\sum_{m,m^{\prime},m_{L},m_{s}}C_{1m,1m^{\prime}}^{Sm_{s}}C_{Sm_{s},Lm_{L}}^{JM}
\epsilon_{n^{\prime}}^{m^{\prime}}\epsilon_{n}^{m}Y_{Lm_{L}}.
\end{eqnarray}

\begin{table}
\centering
\caption{The flavor wave functions of the
$\mathcal{P}-\mathcal{P}$ type molecular states. The corresponding wave functions for $\mathcal{P}^{\ast}-\mathcal{P}^{\ast}$
type just change $\mathcal{P}$ to $\mathcal{P}^{\ast}$ and $\Phi$ to $\Phi^{\ast\ast}$.  \label{Tab:flavor1}}
\begin{tabular}{c>{\footnotesize}cc>{\footnotesize}c}
\hline state&Charm sector &state&Bottom sector\\
\hline
  $\Phi_{s}^{+}$ &$\bar{D}^{0}D_{s}^{+}$&$\Omega_{s}^{+}$&$B^{+}\bar{B}_{s}^{0}$\\

   $\Phi^{+}$ &$\bar{D}^{0}D^{+}$&$\Omega^{+}$&$B^{+}\bar{B}^{0}$\\

   $\Phi_{s}^{0}$ &${D}^{-}D_{s}^{+}$&$\Omega_{s}^{0}$&$B^{0}\bar{B}_{s}^{0}$\\

  $\Phi^{0}$& $\frac{1}{\sqrt{2}}({D}^{0}\bar{D}^{0}-{D}^{-}{D}^{+})$ &$\Omega^{0}$&$\frac{1}{\sqrt{2}}({B}^{0}\bar{B}^{0}-{B}^{-}{B}^{+})$\\

  $\bar{\Phi}_{s}^{0}$ &$D_{s}^{-}D^{+}$&$\bar{\Omega}_{s}^{0}$&$B_{s}^{0}\bar{B}^{0}$\\

    $\Phi^{-}$ &${D}^{-}D^{0}$&$\Omega^{-}$&$B^{0}{B}^{-}$\\

  $\Phi_{s}^{-}$ &${D}_{s}^{-}D^{0}$&$\Omega_{s}^{-}$&$B_{s}^{0}{B}^{-}$\\

  $\Phi_{8}^{0}$& $\frac{1}{\sqrt{2}}({D}^{0}\bar{D}^{0}+{D}^{-}{D}^{+})$ &$\Omega_{8}^{0}$&$\frac{1}{\sqrt{2}}({B}^{0}\bar{B}^{0}+{B}^{-}{B}^{+})$\\

     $\Phi_{s1}^{0}$ &${D}_{s}^{-}D_{s}^{+}$&$\Omega_{s1}^{0}$&$B_{s}^{0}\bar{B}_{s}^{0}$\\

  \hline
  \end{tabular}
\end{table}

Here, we adopt the same convention for the naming of possible molecular states  as used in Ref. \cite{Liu:2008tn}. For the $\mathcal{P}-\mathcal{P}$ type molecular states, we use $\Phi$ and $\Omega$ to indicate the possible molecular state in the charm and bottom sectors, respectively. The detailed formulas of the flavor wave functions can be found in Table \ref{Tab:flavor1}. In the same way, the flavor functions of the $\mathcal{P}^\ast-\mathcal{P}^\ast$ type molecular states can be constructed with $\Phi^{\ast \ast}$ and $\Omega^{\ast \ast}$ as the name of the states in the charm and bottom sectors, respectively. The $\Phi^{\ast}$ and $\Omega^{\ast}$ denote $\mathcal{P}-\mathcal{P}^{\ast}$ type
systems for the charm and bottom sectors, respectively. The detailed formulas of the flavor wave functions of $\mathcal{P}^\ast -\mathcal{P}$ type molecular states are listed in Table \ref{Tab:flavor2}. Here the parameter $c=+1$ and $c=-1$ correspond to the charge parity being negative and positive, respectively. Here we add a hat over the $\Phi^\ast$ and $\Omega^\ast$ to indicate the negative charge parity states.

\begin{table}
\begin{center}
\caption{\small The flavor wave functions of the $\mathcal{P}^\ast -\mathcal{P}$ systems for the charm sector. The corresponding wave functions for the bottom sector can be constructed by replacing the charmed mesons with the corresponding bottom mesons. \label{Tab:flavor2}}
\begin{tabular}{>{\footnotesize}c>{\footnotesize}c>{\footnotesize}c>{\footnotesize}c}
\hline state&Charm sector\\
\hline
$\Phi_{s}^{\ast+}/\widehat{\Phi}_{s}^{\ast+}$&$\frac{1}{\sqrt{2}}(\bar{D}^{\ast0}D_{s}^{+}+c\bar{D}^{0}D_{s}^{\ast+})$\\

$\Phi^{\ast+}/\widehat{\Phi}^{\ast+}$&$\frac{1}{\sqrt{2}}(\bar{D}^{\ast0}D^{+}+c\bar{D}^{0}D^{\ast+})$\\

$\Phi_{s}^{\ast0}/\widehat{\Phi}_{s}^{\ast0}$&$\frac{1}{\sqrt{2}}({D}^{\ast-}D_{s}^{+}+c{D}^{-}D_{s}^{\ast+})$\\

$\Phi^{\ast0}/\widehat{\Phi}^{\ast0}$&
$\frac{1}{2}[({D}^{0}\bar{D}^{\ast0}-{D}^{\ast-}{D}^{+})+c({D}^{\ast0}\bar{D}^{0}-{D}^{-}{D}^{\ast+})]$\\

$\bar{\Phi}_{s}^{\ast0}/\widehat{\bar{\Phi}}_{s}^{\ast0}$
&$\frac{1}{\sqrt{2}}(D_{s}^{\ast-}D^{+}+cD_{s}^{-}D^{\ast+})$\\

$\Phi^{\ast-}/\widehat{\Phi}^{\ast-}$
&$\frac{1}{\sqrt{2}}({D}^{\ast-}D^{0}+c{D}^{-}D^{\ast0})$&\\

$\Phi_{s}^{\ast-}/\widehat{\Phi}_{s}^{\ast-}$
&$\frac{1}{\sqrt{2}}({D}_{s}^{\ast-}D^{0}+c{D}_{s}^{-}D^{\ast0})$\\

$\Phi_{8}^{\ast0}/\widehat{\Phi}_{8}^{\ast0}$&
$\frac{1}{2}[({D}^{0}\bar{D}^{\ast0}+{D}^{\ast-}{D}^{+})+c({D}^{\ast0}\bar{D}^{0}+{D}^{-}{D}^{\ast+})]$
\\

$\Phi_{s1}^{\ast0}/\widehat{\Phi}_{s1}^{\ast0}$
&$\frac{1}{\sqrt{2}}({D}_{s}^{\ast-}D_{s}^{+}+c{D}_{s}^{-}D_{s}^{\ast+})$
\\
\hline
  \end{tabular}
\end{center}
\end{table}

\subsection{Potential of the $\mathcal{P}^{(\ast)}-\mathcal{P}^{(\ast)}$ system}

\begin{figure}
\begin{tabular}{cc}
\begin{minipage}[t]{0.4\linewidth}
\begin{overpic}[scale=.7]{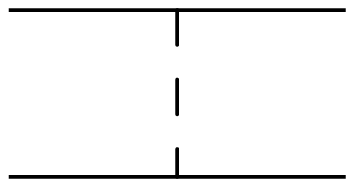}
\put(101,-4){$\mathcal{P}$}
\put(-12,-4){$\mathcal{P}$}
\put(-12,48){$\mathcal{P}$}
\put(101,48){$\mathcal{P}$}
\put(55,24){$\sigma, \mathbb{V}$}
\put(45,-15){(a)}
\end{overpic}
\end{minipage}
&
\begin{minipage}[t]{0.4\linewidth}
\begin{overpic}[scale=.7]{for12.eps}
\put(101,-4){$\mathcal{P}$}
\put(-12,-4){$\mathcal{P}$}
\put(-12,48){$\mathcal{P}^\ast$}
\put(101,48){$\mathcal{P}^\ast$}
\put(55,24){$\sigma, \mathbb{V}$}
\put(45,-15){(b)}
\end{overpic}
\end{minipage}
 \\
 &\\
 &\\
\begin{minipage}[t]{0.4\linewidth}
\begin{overpic}[scale=.7]{for12.eps}
\put(101,-4){$\mathcal{P}^\ast$}
\put(-12,-4){$\mathcal{P}$}
\put(-12,48){$\mathcal{P}^\ast$}
\put(101,48){$\mathcal{P}$}
\put(55,24){$\mathbb{P}, \mathbb{V}$}
\put(45,-15){(c)}
\end{overpic}
\end{minipage}
&
\begin{minipage}[t]{0.4\linewidth}
\begin{overpic}[scale=.7]{for12.eps}
\put(101,-4){$\mathcal{P}^\ast$}
\put(-12,-4){$\mathcal{P}^\ast$}
\put(-12,48){$\mathcal{P}^\ast$}
\put(101,48){$\mathcal{P}^\ast$}
\put(55,24){$\sigma, \mathbb{P},\mathbb{V}$}
\put(45,-15){(d)}
\end{overpic}
\end{minipage}
\end{tabular}
\caption{Feynman diagrams describing  $\mathcal{P}^{(\ast)}-\mathcal{P}^{(\ast)}$ scattering in the one-boson-exchange model. Here $\mathbb{V}$ and $\mathbb{P}$ indicate the light vector and pseudoscalar mesons, respectively.\label{Fig:OBE}}
\end{figure}

The potential of the $\mathcal{P}^{(\ast)}-\mathcal{P}^{(\ast)}$ system can be estimated from the amplitude of the $\mathcal{P}^{(\ast)} \mathcal{P}^{(\ast)} \to \mathcal{P}^{(\ast)}\mathcal{P}^{(\ast)}$ process. Here, we adopt the one-boson-exchange model, where the interaction can be realized by exchanging a light boson as shown in Fig. \ref{Fig:OBE}. The interactions of the heavy-light mesons and light mesons are described by the effective Lagrangian, which are constructed in heavy quark limit and chiral symmetry. The concrete Lagrangians are \cite{Casalbuoni:1996pg, Cheng:1992xi, Yan:1992gz, Wise:1992hn, Burdman:1992gh, Falk:1992cx},
\begin{eqnarray}
\mathcal{L}_{\mathcal{P}^{(\ast)} \mathcal{P}^{(\ast)} \sigma} &=& -2g_{s}\mathcal{P}_{a}\mathcal{P}_{a}^{\dag }\sigma + 2g_{s}\mathcal{P}_{a}^{\ast }\mathcal{P}_{a}^{\ast \dag }\sigma\\
\mathcal{L}_{\mathcal{P}^{(\ast)} \mathcal{P}^{(\ast)} \mathbb{V}} &=& -\sqrt{2}\beta
g_{V}\mathcal{P}_{b} \mathcal{P}_{a}^{\dag }v\mathbb{V}_{ba} -2\sqrt{2}\lambda g_{V}v^{\lambda }\varepsilon _{\lambda \mu
\alpha \beta }(\mathcal{P}_{b}\mathcal{P}_{a}^{\ast \mu \dag
}\nonumber\\&&+\mathcal{P}_{b}^{\ast \mu }\mathcal{P}_{a}^{\dag
})(\partial ^{\alpha }\mathbb{V}^{\beta })_{ba} +\sqrt{2}\beta g_{V}\mathcal{P}_{b}^{\ast }\mathcal{P}_{a}^{\ast
\dag }v\mathbb{V}_{ba}\nonumber\\&&-i2\sqrt{2}\lambda
g_{V}\mathcal{P}_{b}^{\ast \mu }\mathcal{P}_{a}^{\ast \nu \dag
}(\partial _{\mu }\mathbb{V}_{\nu }-\partial _{\nu }\mathbb{V}_{\mu })_{ba}\\
\mathcal{L}_{\mathcal{P}^{(\ast)} \mathcal{P}^{(\ast)} \mathbb{P}} &=&-\frac{2g}{f_{\pi }}(\mathcal{P}_{b}\mathcal{P}_{a\lambda
}^{\ast \dag }+\mathcal{P}_{b\lambda }^{\ast
}\mathcal{P}_{a}^{\dag })\partial ^{\lambda }\mathbb{P}_{ba}
\nonumber \\
&&-i\frac{2g}{f_{\pi }}\varepsilon _{\alpha \mu \nu \lambda
}v^{\alpha }\mathcal{P}_{b}^{\ast \mu }\mathcal{P}_{a}^{\ast
\lambda \dag }\partial ^{\nu }\mathbb{P} _{ba}
\end{eqnarray}
where $g_s =g_\pi/(2\sqrt{6})$, $g_\pi=3.73$, $f_\pi=132 \ \mathrm{MeV}$, $\beta=0.59$, $g_V=5.8$ and $\lambda=0.56\ \mathrm{GeV}^{-1}$, respectively \cite{Casalbuoni:1996pg, Isola:2003fh, Bando:1987br}. The gauge coupling $g=0.59$ is estimated from the experimental width of $D^{\ast+}$ with the assumption that the $D^{\ast +}$ dominantly decays into $D\pi$ \cite{Isola:2003fh}. The light pseudoscalar and vector meson matrices in the above effective Lagrangians are defined as
\begin{eqnarray}
\mathbb{P}=%
\left(\begin{matrix} \frac{\pi^{0}}{\sqrt{2}}+\alpha\eta +\beta
\eta^{\prime } & \pi^{+} & K^{+} \\
\pi ^{-} & -\frac{\pi ^{0}}{\sqrt{2}}+\alpha \eta +\beta \eta
^{\prime }
& K^{0} \\
K^{-} & \overline{K}^{0} & \gamma \eta +\delta \eta ^{\prime }%
\end{matrix}\right)%
\end{eqnarray}
\begin{equation}
\mathbb{V}=%
\left(\begin{matrix} \frac{\rho
^{0}}{\sqrt{2}}+\frac{\omega}{\sqrt{2}} & \rho ^{+} &
K^{\ast +} \\
\rho ^{-} & -\frac{\rho ^{0}}{\sqrt{2}}+\frac{\omega}{\sqrt{2}} &
K^{\ast 0}
\\
K^{\ast -} & \bar{K}^{\ast 0} & \phi%
\end{matrix}\right)%
\end{equation}
where the parameters $\alpha$, $\beta$, $\gamma$ and $\delta$ can be related to the mixing angle $\theta$ by
\begin{eqnarray}
\alpha &=&(\cos
\theta-\sqrt{2}\sin\theta)/\sqrt{6} \nonumber\\
\beta &=&(\sin
\theta+\sqrt{2}\cos \theta)/\sqrt{6} \nonumber\\
\gamma &=&(-2\cos
\theta-\sqrt{2}\sin \theta)/\sqrt{6} \nonumber\\
\delta &=&(-2\sin
\theta +\sqrt{2}\cos \theta )/\sqrt{6}
\end{eqnarray}
and in the present calculations, we use $\theta =-19.1^{\circ }$ \cite{Coffman:1988ve,Jousset:1988ni}.

With the above preparations, we can get the elastic scattering amplitudes corresponding to the diagrams in Fig. \ref{Fig:OBE}.
In general, the scattering amplitude $i\mathcal{M}(J,J_{z})$ can be related to the interaction potential in
 momentum space in terms of the Breit approximation by  \cite{Sun:2011uh,Sun:2012zzd,Liu:2008fh}
\begin{eqnarray}
V(q)=-\frac{\mathcal{M}}{\sqrt{\amalg_{i}2M_{i}\amalg_{f}2M_{f}}}.
\end{eqnarray}
Here, all the involved particles are mesons, so to depict the inner structure effect of the mesons, a monopole type form factor is introduced, which is \cite{Tornqvist:1993vu,Tornqvist:1993ng,Locher:1993cc,Li:1996yn}
\begin{eqnarray}
F(q)=\frac{\Lambda^{2}-m^2}{\Lambda^{2}-q^2} \label{Eq:FF}
\end{eqnarray}
where q is the four-momentum of the exchanged meson. $\Lambda$ is a model parameter, which should be of order 1 GeV.  The effective potential in coordinate space is the Fourier transformation of that in momentum space, and is
\begin{eqnarray}
V(r)=\int\frac{d^{3}p}{(2\pi)^{3}}e^{iqr}V(q)F(q)^{2}.
\end{eqnarray}

In the following, we take the charm sector as an example to show the potentials of the $\mathcal{P}-\mathcal{P}$, $\mathcal{P}^\ast-\mathcal{P}$ and $\mathcal{P}^\ast-\mathcal{P}^\ast$ systems one by one.

\subsubsection{$\mathcal{P}-\mathcal{P}$  type}
As shown in Fig. \ref{Fig:OBE}(a), the interactions between $\mathcal{P}-\mathcal{P}$ can be realized by exchanging a $\sigma$ meson or a vector meson. The corresponding potentials are
\begin{eqnarray}
V_{\sigma }^a(r) &=&-g_{s}^{2}Y(\Lambda ,m_{\sigma },r), \nonumber\\
V_{\mathbb{V}}^a (r) &=&-\frac{1}{2}\beta ^{2}g_{V}^{2}Y(\Lambda ,m_{\mathbb{V}},r),
\end{eqnarray}
respectively, and the the concrete form of the $Y(\Lambda ,m,r)$ is
\begin{eqnarray}
Y(\Lambda ,m,r) &=&\frac{1}{4\pi r}(e^{-m r}-e^{-\Lambda r})-\frac{%
\Lambda ^{2}-m^{2}}{8\pi \Lambda }e^{-\Lambda r}.
\end{eqnarray}
For the $\Phi_s$ and $\Omega_s$ system, their components are $D_s^{-}-D/D_s^+ -\bar{D}$ and $B_s^0-\bar{B}/\bar{B}_s^0-B$, respectively. No proper vector meson can be exchanged in these systems due to the ideal mixing of the $\omega-\phi$. Thus, in the present model, the potential of the $\Phi_s$ and $\Omega_s$ system is zero. The concrete potentials of the $\Phi^\pm$, $\Phi_8^0$ and $\Phi_{s1}^0$ system are
\begin{eqnarray}
V_{\Phi ^{\pm }}&=&-\frac{1}{2}V_{\rho }^a (r)+\frac{1}{2}V_{\omega }^a (r)+V_{\sigma }^a (r)\nonumber\\
V_{\Phi _{8}^{0}}(r)&=&\frac{3}{2}V^a_{\rho }(r)+\frac{1}{2}V^a_{\omega }(r)+V^a_{\sigma }(r)\nonumber\\
V^a_{\Phi _{s1}^{0}}(r)&=&V^a_{\phi }(r)
\label{Eq:PotentialA}
\end{eqnarray}
respectively.

\subsubsection{$\mathcal{P}^{\ast }-\mathcal{P}$ type}
For the $\mathcal{P}^\ast-\mathcal{P}$ system, there exist two
kind of diagrams, the direct diagram and cross diagram, which are
presented in Figs. \ref{Fig:OBE}(b) and \ref{Fig:OBE}(c),
respectively. For the direct diagrams, the exchanged mesons are
$\sigma$ and vector mesons, and the corresponding potentials
are,
\begin{eqnarray}
V_\sigma^{b} &=&-g_{s}^{2}(\varepsilon _{1}\cdot\varepsilon
_{3}^{\dag}) Y(\Lambda,m_{\sigma},r) \nonumber\\
V_\mathbb{V}^{b} &=&-\frac{1}{2}\beta^{2}g_{V}^{2}(\varepsilon
_{1}\cdot\varepsilon _{3}^{\dag}) Y(\Lambda,m_{\mathbb{V}},r),
\end{eqnarray}
respectively.

For the cross diagram, the exchanged light mesons are pseudoscalar and vector mesons. The potentials are
\begin{eqnarray}
V_{\mathbb{P} }^{c}(r)&=&-\frac{g^{2}}{f_{\pi
}^{2}}(\frac{1}{3}(\varepsilon
_{1}\varepsilon _{4}^{\dag })\nabla ^{2}Y(\Lambda
_{0},m_{0},r) \nonumber \\ &&+\frac{1}{3}S(%
\widehat{r},\varepsilon _{1},\varepsilon _{4}^{\dag })r\frac{\partial }{\partial r}\frac{1}{r}\frac{
\partial }{\partial r}Y(\Lambda _{0},m_{0},r),\nonumber\\
V_{\mathbb{V}}^{c}(r)&=&-2\lambda
^{2}g_{V}^{2}(\frac{2}{3}(\varepsilon
_{1}\varepsilon _{4}^{\dag })\nabla ^{2}Y(\Lambda _{0},m_{0},r)\nonumber \\&&-\frac{1}{3}S(%
\widehat{r},\varepsilon _{1},\varepsilon _{4}^{\dag })r\frac{\partial }{%
\partial r}\frac{1}{r}\frac{\partial }{\partial r}Y(\Lambda _{0},m_{0},r)),
\end{eqnarray}
respectively, where $\Lambda_0=\sqrt{\Lambda^2-\Delta^2}$, $m_0=\sqrt{|\Delta^2-m|}$ with $\Delta$ and $m$ being the mass difference of the $\mathcal{P}^\ast$ and $\mathcal{P}$ and the mass of the exchanged meson, respectively.

Here one should notice that the mass splitting of the $D$ and $D^\ast$ meson could be larger than the mass of the $\pi$, thus the exchanged pion meson could be on shell, thus the $Y$ function for the pion exchange is different from other pseudoscalar or vector meson exchange processes, and the $Y$ function for the $\pi$ exchange process is,
\begin{eqnarray}
Y(\Lambda _{0},m_{0},r)_\pi&=&\frac{1}{4\pi r} \big(-e^{-\Lambda
_{0}r}-\frac{r(\Lambda _{0}^{2}+m_{0}^{2})}{2\Lambda
_{0}}e^{-\Lambda _{0}r}+\cos (m_{0}r)\big)   \nonumber \\
\end{eqnarray}

The concrete potentials for the $\Phi_s^\ast$, $\Phi^\ast$,
$\Phi_8^{\ast 0}$ and $\Phi_{s1}^{\ast 0}$ are
\begin{eqnarray}
V_{\Phi_s^\ast}(r) &=&-c\cdot \alpha \gamma V_{\eta
}^{c}(r)-c\cdot \beta \delta V_{\eta ^{^{\prime
}}}^{c}(r) ,\nonumber\\
V_{\Phi^\ast}(r) &=&V_{\sigma }^{b}(r)-\frac{c}{2}V_{\pi
}^{c}(r)+c\cdot \alpha ^{2}V_{\eta }^{c}(r)+c\cdot \beta
^{2}V_{\eta ^{^{\prime }}}^{c}(r) \nonumber \\
&&-\frac{1}{2}( c\cdot V_{\rho }^{c}(r)+V_{\rho
}^{b}(r))+\frac{1}{2}(c\cdot V_{\omega }^{c}(r)+V_{\omega
}^{b}(r)),\nonumber\\
V_{\Phi_{8}^{0\ast}}(r) &=&V_{\sigma }^{b}(r)+3\cdot
\frac{c}{2}V_{\pi }^{c}(r)+c\cdot \alpha ^{2}V_{\eta
}^{c}(r)+c\cdot \beta ^{2}V_{\eta ^{^{\prime
}}}^{c}(r)\nonumber\\
&&+\frac{3}{2}(c\cdot V_{\rho }^{c}(r)+V_{\rho }^{b}(r))
+\frac{1}{2}(c\cdot V_{\omega }^{c}(r)+V_{\omega
}^{b}(r)) ,\nonumber\\
V_{\Phi _{s1}^{\ast 0}}(r) &=&c\cdot \gamma ^{2}V_{\eta
}^{c}(r)+c\cdot \delta ^{2}V_{\eta ^{^{\prime }}}^{c}(r)+(c\cdot
V_{\phi }^{c}(r)+V_{\phi }^{b}(r)) , \nonumber\\\label{Eq:PotentialB}
\end{eqnarray}
respectively. In the above potential,  notice that there
exist two factors related to $\epsilon_i$, which is the
polarization vector of the involved vector mesons. In the subspace
formed by $|^3S_1\rangle$ and $|^3D_1\rangle$, the factor
$\varepsilon _{1}\varepsilon _{3}^{\dag }$ and
$S(\widehat{r},\varepsilon _{1},\varepsilon _{3}^{\dag })$ can
be expressed in matrix form as
\begin{eqnarray}
\varepsilon _{1}\varepsilon _{3}^{\dag }(\varepsilon _{1}\varepsilon _{4}^{\dag }) &=&%
\left(\begin{matrix}
1 & 0 \\
0 & 1%
\end{matrix}\right),
\\
S(\widehat{r},\varepsilon _{1},\varepsilon _{3}^{\dag }) &=&%
\left(\begin{matrix}
0 & -\sqrt{2} \\
-\sqrt{2} & 1%
\end{matrix}\right)
\end{eqnarray}
respectively.

\subsubsection{$\mathcal{P}^{\ast}$-$\mathcal{P}^{\ast }$ type}
For the $\mathcal{P}^\ast-\mathcal{P}^\ast$ system, the exchanged mesons can be $\sigma$, pseudoscalar and vector mesons as shown in Fig. \ref{Fig:OBE}(d). The corresponding potentials are
\begin{eqnarray}
V^{d}_{\sigma} &=&-g_{s}^{2}(\varepsilon _{1}\times \varepsilon
_{3}^{\dag
})\cdot (\varepsilon _{2}\times \varepsilon _{4}^{\dag })Y(\Lambda,m_{\sigma},r)\nonumber\\
V^{d}_{\mathbb{P}}(r)&=&-\frac{g^{2}}{f_{\pi
}^{2}}(\frac{1}{3}(\varepsilon _{1}\times \varepsilon _{3}^{\dag
})\cdot (\varepsilon _{2}\times \varepsilon _{4}^{\dag })\nabla
^{2}Y(\Lambda
,m_{\mathbb{P}},r))\nonumber\\&&+\frac{1}{3}S(\widehat{r},\varepsilon _{1}\times
\varepsilon _{3}^{\dag },\varepsilon _{2}\times \varepsilon
_{4}^{\dag })r\frac{\partial }{\partial
r}\frac{1}{r}\frac{\partial }{\partial r})Y(\Lambda
,m_{\mathbb{P}},r))\nonumber\\
V^d_{\mathbb{V}}(r) &=&-\frac{1}{2}\beta ^{2}g_{V}^{2}(\varepsilon
_{1}\varepsilon _{3}^{\dag })(\varepsilon _{2}\varepsilon
_{4}^{\dag })Y(\Lambda ,m_{\mathbb{V}},r) \nonumber \\&&-2\lambda
^{2}g_{V}^{2}(\frac{1}{3}(\varepsilon _{1}\times \varepsilon
_{3}^{\dag })(\varepsilon _{2}\times \varepsilon _{4}^{\dag
})\nabla ^{2}Y(\Lambda ,m_{\mathbb{V}},r)\nonumber\\
&& -\frac{2}{3}S(\widehat{r},\varepsilon _{1}\times \varepsilon
_{3}^{\dag },\varepsilon _{2}\times \varepsilon _{4}^{\dag
})r\frac{\partial }{\partial r}\frac{1}{r}\frac{\partial
}{\partial r})Y(\Lambda ,m_{\mathbb{V}},r)),\nonumber \\
\end{eqnarray}
respectively.

The total potentials of the $\Phi_s^{\ast \ast}$, $\Phi_{s1}^{\ast\ast}$, $\Phi^{\ast \ast}$ and $\Phi_8^{0\ast \ast}$ systems are
\begin{eqnarray}
V_{\Phi_s^{\ast \ast}}(r)&=&\alpha \gamma V_{\eta }^{d}(r)+\beta
\delta V_{\eta
^{^{\prime }}}^{d}(r) \nonumber\\
V_{\Phi_{s1}^{0\ast \ast}}(r) &=& \gamma
^{2}V_{\eta}^{d}(r)+\delta
^{2}V_{\eta ^{\prime }}^{d}(r)+V_{\phi}^{d}(r)\nonumber\\
V_{\Phi^{\ast \ast}}(r)&=& -\frac{1}{2}V_{\pi}^{d}(r)+\alpha
^{2}V_{\eta }^{d}(r)+\beta ^{2}V_{\eta ^{^{\prime }}}^{d}(r)
-\frac{1}{2}V_{\rho }^{d}(r)+\frac{1}{2}V_{\omega}^{d}(r) \nonumber\\
V_{\Phi_{8}^{0 \ast \ast }}(r)&=&\frac{3}{2}V_{\pi}^{d}(r)+\alpha
^{2}V_{\eta}^{d}(r)+\beta ^{2}V_{\eta ^{^{\prime
}}}^{d}(r)+\frac{3}{2}
V_{\rho}^{d}(r) +\frac{1}{2}V_{\omega}^{d}(r),\nonumber\\
\label{Eq:PotentialC}
\end{eqnarray}
respectively. The factor related to the polarization vectors of the involved vector mesons can be expressed in matrix form as
\begin{eqnarray}
(\varepsilon _{1}\varepsilon _{3}^{\dag })(\varepsilon
_{2}\varepsilon
_{4}^{\dag }) &= &%
\left\{
  \begin{array}{ll}
    \left(\begin{matrix}
1 & 0 \\
0 & 1%
\end{matrix}\right), & \hbox{J=0}
\\
\left(\begin{matrix}
1 & 0 & 0 \\
0 & 1 & 0 \\
0 & 0 & 1%
\end{matrix}\right), & \hbox{J=1}\\
\left(\begin{matrix}
1 & 0 & 0 & 0 \\
0 & 1 & 0 & 0 \\
0 & 0 & 1 & 0 \\
0 & 0 & 0 & 1%
\end{matrix}\right), & \hbox{J=2}
  \end{array}
\right., \nonumber
\end{eqnarray}
\begin{eqnarray}
(\varepsilon _{1}\times \varepsilon _{3}^{\dag })(\varepsilon
_{2}\times
\varepsilon _{4}^{\dag })=
\left\{
  \begin{array}{ll}
\left(\begin{matrix}
2 & 0 \\
0 & -1%
\end{matrix}\right), & \hbox{J=0}
\\
\left(\begin{matrix}
1 & 0 & 0 \\
0 & 1 & 0 \\
0 & 0 & -1%
\end{matrix}\right), & \hbox{J=1}\\
\left(\begin{matrix}
-1 & 0 & 0 & 0 \\
0 & 2 & 0 & 0 \\
0 & 0 & 1 & 0 \\
0 & 0 & 0 & -1%
\end{matrix}\right), & \hbox{J=2}
  \end{array}
\right., \nonumber
\end{eqnarray}

\begin{eqnarray}
S(\widehat{r},\varepsilon _{1}\times \varepsilon _{3}^{\dag
},\varepsilon
_{2}\times \varepsilon _{4}^{\dag }) =%
\left\{
  \begin{array}{ll}
\left(\begin{matrix}
0 & \sqrt{2} \\
\sqrt{2} & 2
\end{matrix}\right), & \hbox{J=0}
\\
\left(\begin{matrix}
0 & -\sqrt{2} & 0 \\
-\sqrt{2} & 1 & 0 \\
0 & 0 & 1%
\end{matrix}\right), & \hbox{J=1}\\
\left(\begin{matrix}
0 & \sqrt{\frac{2}{5}} & 0 & -\sqrt{\frac{14}{5}} \\
\sqrt{\frac{2}{5}} & 0 & 0 & -\frac{2}{\sqrt{7}} \\
0 & 0 & -1 & 0 \\
-\sqrt{\frac{14}{5}} & -\frac{2}{\sqrt{7}} & 0 & -\frac{3}{7}%
\end{matrix}\right), & \hbox{J=2}
  \end{array}
\right.,\nonumber
\end{eqnarray}
respectively.

The matrix forms of the kinetic terms for $\mathcal{P}-\mathcal{P}$, $\mathcal{P}^\ast-\mathcal{P}/ \mathcal{P}^\ast- \mathcal{P}^\ast(J=0)$, $\mathcal{P}^\ast- \mathcal{P}^\ast(J=1)$,$\mathcal{P}^\ast- \mathcal{P}^\ast(J=2)$ are \begin{eqnarray}
K&=&diag(-\frac{\bigtriangleup}{2\mu})\\
K&=&diag(-\frac{\bigtriangleup}{2\mu},-\frac{\bigtriangleup_{1}}{2\mu})\\
K&=&diag(-\frac{\bigtriangleup}{2\mu},-\frac{\bigtriangleup_{1}}{2\mu},-\frac{\bigtriangleup_{1}}{2\mu})\\
K&=&diag(-\frac{\bigtriangleup}{2\mu},-\frac{\bigtriangleup_{1}}{2\mu}, -\frac{\bigtriangleup_{1}}{2\mu}, -\frac{\bigtriangleup_{1}}{2\mu})
\end{eqnarray}
respectively. Here, $\bigtriangleup=\frac{1}{r^{2}}\frac{\partial}{\partial
r}r^{2}\frac{\partial}{\partial
r}$, $\bigtriangleup_{1}=\bigtriangleup-\frac{6}{r^{2}}$ and $\mu$ is the reduced mass of the considered system. With the potentials listed in Eqs. (\ref{Eq:PotentialA})-(\ref{Eq:PotentialC}) and the above kinetic terms, one can get bound energies and wave functions if there exist bound states by solving the corresponding Schr$\ddot{\mathrm{o}}$dinger equation. In the present work, we rely on complex scaling methods to perform the calculations, in which the wave function of the bound state is expanded by the harmonic oscillator wave functions \cite{Moiseyev:1988nm,Guo:2010cpc,Liu:2013lq,Guo:2010zzm, Liu:2012da, Alhaidari:2001AD}.

\section{Numerical results and discussion \label{Sec:3}}
In the OBE model, one additional cutoff $\Lambda$ is introduced in the form factor, which compensates the off-shell effect of the changed light mesons. The value of the $\Lambda$ should be of order 1 GeV and in the present work, we search the bound state solutions of different systems with $\Lambda$ less than 3 GeV, which is a reasonable cutoff for light meson exchange processes. In the following, we will present the numerical results of the three types of system separately.

\begin{table}[!h]
\centering
\begin{tabular}{c|cccc}
\hline\hline
&state& $~~\Lambda$ (GeV) ~~&~~$E$ (MeV)~~&~~$r_{\mathrm{RMS}}$ (fm)~~\\
\hline
\multirow{3}{5mm}{\rotatebox{90}{ Charm}}& \multirow{3}{0.2cm}{$\Phi_{8}^{0}$}
    &1.50  &-0.43   &2.27 \\
    && 1.60 & -3.65  & 1.76\\
    && 1.70 & -8.34  & 1.36 \\
 \hline
\multirow{6}{5mm}{\rotatebox{90}{ Bottom}} &\multirow{3}{0.2cm}{$\Omega_{8}^{0}$}
     &1.10  &-1.15   &1.85\\
    && 1.20 & -8.33  &0.97\\
    && 1.30 & -20.81 &0.70 \\
\cline{2-5}
&\multirow{3}{0.5cm}{$\Omega_{s1}^{0}$}
     &1.90  &-0.26   &2.01 \\
    &&2.10  & -3.69  & 1.10   \\
    && 2.30 & -9.67  & 0.77   \\
\hline\hline
\end{tabular}
\caption{The binding energy and the root-means-square radius of the $\mathcal{P}-\mathcal{P}$ type molecular state depending on the cutoff $\Lambda$.
\label{Tab:p-ptype}}
\end{table}

\begin{table}[!h]
\centering
\begin{tabular}{c|c|c|c}
\hline\hline
 System &Molecule state &Present work & Ref. \cite{Liu:2008mi} \\ \hline
\multirow{2}{5mm}{$D\bar{D}$}       &$\Phi_{8}^{0}$  &$\surd$   &$\ast$ \\
       &\multirow{1}{3.5mm}{$\Phi^{0(\pm)}$ }&$\otimes$   &$\otimes$\\
      \hline
$D_s  \bar{D}$  &\multirow{1}{3.5mm}{$\Phi_{s}^{\pm}$}  &$\otimes$  &$\otimes$
        \\ \hline
$D_s^+ D_s^-$ & \multirow{1}{3.5mm}{$\Phi_{s1}^{0}$}  &$\otimes$    &?
       \\ \hline
\multirow{2}{5mm}{$B\bar{B}$}  &      \multirow{1}{3.5mm}{$\Omega_{8}^{0}$}  &$\surd$   &$\surd$\\
  &     \multirow{1}{3.5mm}{$\Omega^{0(\pm)}$}  &$\otimes$   &?\\
               \hline
$B_s^0\bar{B}$ &$\Omega_{s}^{\pm}$  &$\otimes$   &$\otimes$
        \\ \hline
 $B_s^0 \bar{B}_s^0$& \multirow{1}{3.5mm}{$\Omega_{s1}^{0}$}  &$\surd$   &$\ast$\\
 \hline\hline
\end{tabular}
\caption{{ Summary of possible bound states for
$\mathcal{P}-\mathcal{P} $ type. Here, we also compare our results with the estimations from the chiral and extended chiral $SU(3)$ quark model \cite{Liu:2008mi}. The symbols $\surd$, $\otimes$ and $?$ indicate that this bound
state must, must not or maybe exists, respectively. The symbol $\ast$  means this bound state does not exist in the chiral $SU(3)$ quark model while it is possible or not excluded in the extended chiral $SU(3)$ quark model.  }\label{Tab:sum1}}
\end{table}

\subsection{$\mathcal{P}-\mathcal{P}$ type}

In the $\mathcal{P}-\mathcal{P}$ type system, we have not observed
any near threshold states which may correspond to a
$\mathcal{P}-\mathcal{P}$ type molecular state. However, in the
present calculations, we find one bound state solution in the charm
sector, the $\Phi_8^{0}$ state with $I(J^{PC})=0(0^{++})$. As
shown in Table \ref{Tab:p-ptype}, when the cutoff increases from
1.5 GeV to 1.7 GeV, the binding energy varies from less than 1 MeV
to nearly 10 MeV, which corresponds to the mass of the $\Phi_8^0$
decreasing from $3738$ MeV to $3729$ MeV. In addition,  in this
cutoff range, the root-mean-square (RMS) radius of the system
decrease from 2.27 fm to 1.36 fm, which indicates the $D \bar{D}$
could form a very loosely shallow bound state by the $\sigma$ and
vector meson exchange.

 In the bottom sector, we find two bound state solutions,
$\Omega_8^0$ and $\Omega_{s1}^0$. The $\Omega_8^0$ state
is the bottom correspondence of the $\Phi_8^0$ in the charm
sector. When we vary the cutoff from 1.10 GeV to 1.30 GeV, the
binding energy of the $\Phi_8^0$ increases from about 1 MeV to
more than 20 MeV, while the RMS radius decreases from 1.85 fm to
0.70 fm. Comparing the binding energies and RMS radii of the
$\Phi_8^0$ and $\Omega_8^0$, the $\Omega_8^0$ is a
more compact bound state than $\Phi_8^0$ for the same binding
energy. In the bottom sector, we also find the bound solution of
$\Omega_{s1}^0$, which is a  $I(J^{PC})=0(0^{++}) $state composed
of $B_s^0 \bar{B}_s^0$. Here we should note that the only possible
exchanged meson of the $B_s^0 \bar{B}_s^0$ is the  $\phi$ meson, which
provides an attractive potential.

{In the $\mathcal{P}-\mathcal{P}$ system, the total spin is zero, thus there is no $S-D$ mixing in such a system. Our present results are consistent with those in Ref. \cite{Liu:2008tn}, in which molecular states $\Phi_{8}^{0}$, $\Phi_{s1}^{0}$, $\Omega_{8}^{0}$ and $\Omega_{s1}^{0}$ states were established, while
no bound state solution corresponding to $\Phi^{\pm(0)}$ and $\Omega^{\pm(0)}$  was found.  We summarize the possible bound states of the $\mathcal{P}-\mathcal{P}$ system in Table \ref{Tab:sum1} and compare with estimations in the chiral and extended chiral $SU(3)$ quark model \cite{Liu:2008mi}. From the table, the present calculations in the OBE potential model are almost in line with the estimations in the chiral and extended chiral $SU(3)$ quark model \cite{Liu:2008mi}. However, our calculation can exclude the possibilities  of $\Phi_{s1}^0$ and $\Omega^{0 \pm}$ as molecular states, while the calculation in Ref. \cite{Liu:2008mi} could not fully exclude such possibilities.}

\begin{table}[!h]
\centering
\begin{tabular}{c|cccc}
\hline\hline
&state& $~~\Lambda$ (GeV) ~~&~~$E$ (MeV)~~&~~$r_{\mathrm{RMS}}$ (fm)~~\\
\hline \multirow{12}{5mm}{\rotatebox{90}{ Charm}}&
\multirow{3}{0.2cm}{$\Phi_{8}^{0\ast}$}
 &1.10  &-0.49   &2.26 \\
    &&1.20  &-5.53  &1.58 \\
    &&1.30  &-14.02 &1.16 \\
    \cline{2-5} &\multirow{3}{0.5cm}{$\widehat{\Phi}_{8}^{0\ast}$}
     &1.40 &-2.18 &2.18 \\
    &&1.50   &-14.03   &1.24   \\
    &&1.60   &-37.41   &0.82   \\
 \cline{2-5} &\multirow{3}{0.5cm}{$\Phi_{s1}^{0\ast}$}
     &2.70 &-1.39 &1.88 \\
    &&2.90   &-8.15   &1.17   \\
    &&3.10   &-20.40   &0.82   \\
 \cline{2-5} &\multirow{3}{0.5cm}{$\widehat{\Phi}_{s1}^{0\ast}$}
     &2.30 &-2.21 &1.96 \\
    &&2.40   &-10.67   &1.22   \\
    &&2.50   &-25.97   &0.84   \\
 \hline
\multirow{12}{5mm}{\rotatebox{90}{ Bottom}}
&\multirow{3}{0.2cm}{$\Omega_{8}^{0\ast}$}
     &0.80  &-1.21  &1.81\\
    &&0.90   &-7.76   &1.03\\
    &&1.00   &-22.13  &0.74 \\
\cline{2-5} &\multirow{3}{0.5cm}{$\widehat{\Omega}_{8}^{0\ast}$}
     &1.00  &-0.27  &2.79 \\
    &&1.05   &-2.19  &2.08   \\
    &&1.10   &-6.15  &1.42   \\
\cline{2-5} &\multirow{3}{0.5cm}{$\Omega_{s1}^{0\ast}$}
      &1.80    &-4.88     &1.05\\
      &&1.90    &-10.37    &0.89\\
      &&2.00    &-19.29    &0.66\\
\cline{2-5} &\multirow{3}{0.5cm}{$\widehat{\Omega}_{s1}^{0\ast}$}
      &1.70  &-4.69  &1.20\\
      &&1.80    &-17.74     &0.69\\
      &&1.90    &-40.52     &0.48\\
 \hline\hline
\end{tabular}
\caption{The same as Table \ref{Tab:p-ptype} but for $\mathcal{P}-\mathcal{P}^\ast $ type.
\label{Tab:p-pstype}}
\end{table}

\begin{table}[!h]
\centering
\begin{tabular}{c|c|c|c}
\hline\hline System &Molecule states&Present Work& Ref. \cite{Liu:2008mi} \\ \hline
  \multirow{4}{0.5cm}{$D\bar{D}^{\ast}$}
        &$\Phi_{8}^{0\ast}$            &$\surd$     &$\surd$\\
        &$\widehat{\Phi}_{8}^{0\ast}$  &$\surd$      &?       \\
        & \multirow{1}{0.4cm}{$\Phi^{0(\pm)\ast}$}                &$\otimes$   &$\otimes$\\
        & \multirow{1}{0.4cm}{$\widehat{\Phi}^{0(\pm)\ast}$}      &$\otimes$   &$\otimes$\\
      \hline
  \multirow{2}{0.5cm}{$\bar{D}{D}_{s}^{\ast}$}& \multirow{1}{0.4cm}{$\Phi_{s}^{0(\pm)\ast}$} &$\otimes$&$\otimes$\\
                                        &\multirow{1}{0.4cm}{$\widehat{\Phi}_{s}^{0(\pm)\ast}$}&$\otimes$&$\otimes$
        \\ \hline
   \multirow{2}{0.9cm}{${D}_{s}^+{D}_{s}^{-\ast}$}
        &$\Phi_{s1}^{0\ast}$             &?   &?\\
        &$\widehat{\Phi}_{s1}^{0\ast}$   &?   &?
       \\ \hline
\multirow{4}{0.5cm}{$B\bar{B}^{\ast}$}
             &$\Omega_{8}^{0\ast}$ &$\surd$   &$\surd$\\
                              &$\widehat{\Omega}_{8}^{0\ast}$ &$\surd$   &$\ast$\\
      & \multirow{1}{0.4cm}{$\Omega^{0(\pm)\ast}$}&$\otimes$   &$\otimes$\\
                               & \multirow{1}{0.4cm}{$\widehat{\Omega}^{0(\pm)\ast}$}&$\otimes$   &$\ast$\\
      \hline
  \multirow{2}{0.5cm}{$B\bar{B}_{s}^{0\ast}$}
         &$\Omega_{s}^{0\ast}$               &$\otimes$    &$\otimes$\\
         &\multirow{1}{0.5cm}{$\widehat{\Omega}_{s}^{0(\pm)\ast}$}     &$\otimes$&$\otimes$
              \\  \hline
              \multirow{2}{0.5cm}{$B_{s}^{0}\bar{B}_{s}^{0\ast}$}
         &$\Omega_{s1}^{0\ast}$  &$\surd$    &$\ast$\\
         &$\widehat{\Omega}_{s1}^{0\ast}$ &$\surd$&$\ast$  \\
 \hline\hline
\end{tabular}
\caption{ The same as Table \ref{Tab:sum1} but for the $\mathcal{P}-\mathcal{P}^\ast$ system.  \label{Tab:sum2}}
\end{table}

\begin{table*}[htb]
\centering
\begin{tabular}{c|cccc|cccc|cccc}
\hline\hline
& \multicolumn{4}{c|}{$J=0$} & \multicolumn{4}{c|}{$J=1$} &\multicolumn{4}{c}{$J=2$}\\
\hline
&state& $~~\Lambda$ (GeV) ~~&~~$E$ (MeV)~~&~~$r_{\mathrm{RMS}}$ (fm)~~
& state& $~~\Lambda$ (GeV) ~~&~~$E$ (MeV)~~&~~$r_{\mathrm{RMS}}$ (fm)~~
& state& $~~\Lambda$ (GeV) ~~&~~$E$ (MeV)~~&~~$r_{\mathrm{RMS}}$ (fm)~~\\
\hline
\multirow{6}{5mm}{\rotatebox{90}{ Charm}}
&\multirow{3}{0.2cm}{$\Phi_{8}^{0\ast\ast}$}
     &1.40   &-2.43     &1.84
&\multirow{3}{0.2cm}{$\Phi_{8}^{0\ast\ast}$}
     &1.40  &-4.41  &1.96
&\multirow{3}{0.2cm}{$\Phi_{8}^{0\ast\ast}$}
     &1.10  &-1.49  &2.12 \\
    &&1.50   &-19.25    &0.99
    &&1.50    &-18.53   &1.11
    &&1.20    &-9.56     &1.34 \\
    &&1.60   &-55.67    &0.70
    &&1.60    &-44.51   &0.75
    &&1.30    &-23.52    &0.97\\
    \cline{2-13}
&\multirow{3}{0.5cm}{$\Phi_{s1}^{0\ast\ast}$}
     &2.15    &-1.09  &1.89
&\multirow{3}{0.5cm}{$\Phi_{s1}^{0\ast\ast}$}
     &2.30   &-3.40  &1.86
&\multirow{3}{0.5cm}{$\Phi_{s1}^{0\ast\ast}$}
     &2.30   &-0.14   &2.32 \\
    &&2.19   &-4.21  &1.42
    &&2.40   &-13.07  &1.11
    &&2.40  &-3.88    &1.66  \\
    &&2.23   &-9.05  &1.07
    &&2.50    &-29.86  &0.78
    &&2.50   &-10.76   &1.17  \\
 \hline
\multirow{6}{5mm}{\rotatebox{90}{ Bottom}}
&\multirow{3}{0.2cm}{$\Omega_{8}^{0\ast\ast}$}
     &1.00   &-1.05   &2.04
&\multirow{3}{0.2cm}{$\Omega_{8}^{0\ast\ast}$}
     &1.00   &-0.36   &3.05
&\multirow{3}{0.2cm}{$\Omega_{8}^{0\ast\ast}$}
     &0.80   &-2.76   &1.53  \\
    &&1.10   &-11.91  &1.07
    &&1.10   &-6.32   &1.40
    &&1.20   &-23.65  &0.76\\
    &&1.20   &-39.85  &0.75
    &&0.90   &-11.59  &0.92
    &&1.00   &-30.00  &0.67\\
\cline{2-13}
&\multirow{3}{0.5cm}{$\Omega_{s1}^{0\ast\ast}$}
     &1.60   &-0.54   &1.86
&\multirow{3}{0.5cm}{$\Omega_{s1}^{0\ast\ast}$}
     &1.70   &-5.38   &1.15
&\multirow{3}{0.5cm}{$\Omega_{s1}^{0\ast\ast}$}
     &1.70   &-7.61   &0.96 \\
    &&1.70   &-10.71  &0.76
    &&1.80   &-19.19  &0.67
    &&1.80   &-13.39  &0.77    \\
    &&1.80   &-35.69  &0.51
    &&1.90   &-42.86  &0.47
    &&1.90   &-35.72  &0.56  \\
 \hline\hline
\end{tabular}
\caption{The same as Table \ref{Tab:p-ptype} but for $\mathcal{P}^\ast -\mathcal{P}^\ast$ type \label{Tab:ps-pstype}}
\end{table*}

\begin{table}[!h]
\centering
\begin{tabular}{c|c|c|c|c}
\hline\hline System& Molecule states &J&Present work & Ref. \cite{Liu:2008mi} \\ \hline
  \multirow{6}{0.5cm}{$D^{\ast}\bar{D}^{\ast}$}
        &\multirow{3}{0.5cm}{$\Phi_{8}^{0\ast\ast}$} &0&$\surd$   &?\\
                              &  &1&$\surd$   &?\\
                               &  &2&$\surd$   &$\surd$\\ \cline{2-5}
       &\multirow{3}{0.5cm}{$\Phi^{0(\pm)\ast\ast}$} & 0&$\otimes$   &$\otimes$\\
                               & &1&$\otimes$   &$\otimes$\\
                                &  &2 &$\otimes$   &$\otimes$\\
      \hline
  \multirow{3}{0.5cm}{$\bar{D}^{\ast}{D}_{s}^{\ast}$}& \multirow{3}{0.5cm}{$\Phi_{s}^{0(\pm)\ast\ast}$}& 0&$\otimes$
  &$\otimes$\\
   &&1&$\otimes$&$\otimes$  \\
  &&2&$\otimes$&$\otimes$
        \\ \hline
   \multirow{3}{0.9cm}{${D}_{s}^{+\ast}{D}_{s}^{-\ast}$}
        &\multirow{3}{0.5cm}{$\Phi_{s1}^{0\ast\ast}$}&0 &?   &?\\
        & &1&?&?    \\
        & &2&?&?
       \\ \hline
\multirow{6}{0.5cm}{$B^{\ast}\bar{B}^{\ast}$}
        &\multirow{3}{0.5cm}{$\Omega_{8}^{0\ast\ast}$} &0&$\surd$   &?\\
                              &  &1&$\surd$   &$\ast$\\
                                       &  &2&$\surd$   &$\surd$\\ \cline{2-5}
       &\multirow{3}{0.5cm}{$\Omega^{0(\pm)\ast\ast}$} & 0&$\otimes$   &$\ast$\\
                               & & 1&$\otimes$   &$\ast$\\
                                        &  &2&$\otimes$   &$\otimes$\\
      \hline
  \multirow{3}{0.5cm}{$B^{\ast}\bar{B}_{s}^{0\ast}$}
 &\multirow{3}{0.5cm}{$\Omega_{s}^{0(\pm)\ast\ast}$}&0 &$\otimes$    &$\otimes$\\
         &&1&$\otimes$&$\otimes$   \\
                 &&2&$\otimes$&$\otimes$
              \\  \hline
              \multirow{3}{0.7cm}{$B_{s}^{0\ast}\bar{B}_{s}^{0\ast}$}
 &\multirow{3}{0.5cm}{$\Omega_{s1}^{0\ast\ast}$}&0  &$\surd$    &$\ast$\\
         &&1&$\surd$&$\ast$  \\
                  &&2&$\surd$&$\ast$  \\
 \hline\hline
\end{tabular}
\caption{{The same as Table \ref{Tab:sum1} but for the $\mathcal{P}^\ast -\mathcal{P}^\ast$ system.} \label{Tab:sum3}}
\end{table}

\subsection{$\mathcal{P}-\mathcal{P}^{\ast}$ type}
For the $\mathcal{P}-\mathcal{P}^{\ast}$ type system, both the spin and total angular momentum are $1$ if only the $S$ wave dominant state is considered. In the present work, $S-D$ mixing is considered. The binding energies and RMS radii of the bound state solutions depending on the cutoff $\Lambda$ are presented in Table \ref{Tab:p-pstype}.

In the charm sector, we get four bound state solutions,
$\Phi_8^{0 \ast}$, $\hat{\Phi}_8^{0 \ast}$, $\Phi_{s1}^{0 \ast}$
and $\hat{\Phi}_{s1}^{0 \ast}$,  where
$\Phi_8^{0\ast}$ corresponds to the experimentally observed
$X(3872)$. When $\Lambda=1.10 $ GeV, the binding energy of the
$\Phi_8^{0 \ast}$ is very small, which agrees with the
experimental observation of the $X(3872)$. In this case, the RMS
radius of the $X(3872)$ could reach up to 2 fm. Thus, the
estimation in the present work indicates that the observed
$X(3872)$ is a very loosely shallow  bound state of the $D
\bar{D}^\ast+h.c$, {which is the same as the conclusion in Ref. \cite{Sun:2012zzd}, qualitatively.
However, the binding energy of the $\Phi_8^{0\ast}$ is smaller than the one in Ref. \cite{Sun:2012zzd} with the same cutoff, due to proper consideration of the $\eta-\eta^\prime$ mixing in the present work. In addition, in the present work, the mass splittings of the charged and neutral charmed mesons are not taken into consideration. In Ref. \cite{Li:2012cs}, both the mass splittings of the charmed mesons and the $S-D$ mixing were considered, and the mass and decays of the $X(3872)$ were well reproduced. We find the mass splittings of the neutral and charged mesons strongly affect the decays of the $X(3872)$, while the mass could be well explained both with and without considering such mass splittings with a reasonable cutoff.}

In addition, the partner of the $X(3872)$
with negative $C$ parity is also predicted in our present
calculations. As the strange partner of the $X(3872)$, the state
$\Phi_{s1}^{0 \ast}$ system has bound state solutions when we take
a relative large cutoff, which is about 3 GeV. For the negative
$C$ parity system, $\hat{\Phi_{s1}^{0 \ast }}$, we can also find
the  bound state solution when $\Lambda$ {is} larger
than 2.3 GeV. As listed in Table \ref{Tab:threshold}, $Z_c(3900)$
is also very close to the threshold of the $D\bar{D}^\ast$ with
$I(J^{P})= 1 (1^+)$.{ In the present calculation, however, we do not
find the bound state of the $D\bar{D}^\ast +h.c$ with $I=1$, which
indicates that the present calculation does not support the observed
$Z_c(3900)$ as the $D\bar{D}^\ast$ molecular state. In Ref. \cite{He:2015mja}, the author
carried out a calculation within the Bethe-Salpeter equation approach and found that $Z_{c}(3900)$ could be a resonance state above $D\bar{D}^{\ast}$ threshold rather than a bound state below $D\bar{D}^\ast$ threshold.}

In the bottom sector, there also exist four bound states in our calculations,  $\Omega_8^{0 \ast}$, $\hat{\Omega}_8^{0 \ast}$, $\Omega_{s1}^{0 \ast}$ and $\hat{\Omega}_{s1}^{0 \ast}$. Compared to the charm correspondence of these states, we find that cutoffs in the bottom sector are smaller than those in the charm sector and in addition, the RMS radii of these states are smaller than their correspondences in the charm sector with the same binding energy. The corresponding state of the $Z_b(10610)$ could not be found in our present calculations, which is the same case as the $Z_c(3900)$.
{In Ref. \cite{Sun:2012zzd}, the estimation in the OBE potential model indicated that the $Z_{b}(10610)$ could be a molecular state of $B\bar{B}^{\ast}$, which is different from our present calculation. The main reason for such a difference is that $\eta-\eta^{\prime}$ mixing is considered in the present work, which increases repulsive interaction for the isospin triplet. In addition, the authors in Ref. \cite{Ohkoda:2011vj} indicated that the $Z_b(10610)$ could be a $B^\ast \bar{B} +h.c$ molecular state, which is different from our present calculation. In Ref. \cite{Ohkoda:2011vj}, the authors considered  $B^\ast \bar{B} -B^\ast \bar{B}^\ast$ mixing but only included the potentials induced by $\pi$, $\rho$ and $\omega$ exchange, which may be the reason for the different conclusions drawn from our present calculation.

In Table \ref{Tab:sum2}, we summarize our calculation for the $\mathcal{P}-\mathcal{P}^\ast$ system and compare with the chiral and extended chiral $SU(3)$ quark model \cite{Liu:2008mi}. Our estimations in the OBE quark model are consistent with those in Ref. \cite{Liu:2008mi} except for the $\hat{\Phi}^{\ast 0}$. In this work, we find a bound state solution for $\hat{\Omega}^{\ast 0}$, while in Ref. \cite{Liu:2008mi}, their calculation indicated that such a state may exist.}

\subsection{$\mathcal{P}^{\ast}-\mathcal{P}^{\ast}$ type}
For the system composed of two red heavy $S-$wave vector mesons, the total angular
momentum of the system could be 0, 1, and 2 for the $S-$wave
interaction. The binding energies and RMS radii of the possible
bound states depending on the cutoff  are presented in Table
\ref{Tab:ps-pstype}. From our calculations, we can find the bound
states of $\Phi_8^{0 \ast \ast}$ and $\ \Phi_{s1}^{0 \ast \ast}$
for different total angular momenta. However, $Y(4140)$ is about
80 MeV below the threshold of the $D_s^{\ast +} D_s^{\ast-}$,
which is larger than the binding energy of the $\Phi_{s1}^{0 \ast
\ast}$. In addition, the LHCb Collaboration have measured the
$J^{PC}$ quantum numbers of the $Y(4140)$ to be $1^{++}$
\cite{Aaij:2016iza, Aaij:2016nsc}, which is different
{from} $\Phi_{s1}^{0 \ast \ast}$. Thus, the $Y(4140)$ cannot be
a $D_s^{\ast + } D_s^{\ast -}$ molecular state.  When taking both the
$S-D$ mixing and $\eta-\eta^{\prime}$ mixing into
consideration, we do not find the bound state corresponding to the
observed $Z_c(4020)$. The calculation in Ref. \cite{Sun:2012zzd} also indicated that there is no
bound state for isovector states with $J=0,1,2$, and only isoscalar bound
states could be found. Our present calculations are consistent with those in Ref. \cite{Sun:2012zzd}, qualitatively, but the binding energies of the obtained molecular states in the present work are a little
bigger than the corresponding ones  with the same cutoff due to $\eta-\eta^{\prime}$ mixing  \cite{Sun:2012zzd}.

In the bottom sector, we also find two group bound states with different total angular momenta, the $\Omega_{8}^{0\ast \ast}$ and $\Omega_{s1}^{0 \ast \ast}$. Similar to the charm sector, our calculations also do not support the molecular interpretations of the $Z_b(10650)$. Similar to the case of $Z_b(10610)$, the estimation in Ref. \cite{Sun:2012zzd} indicated that $Z_b(10650)$ could be a bound state composed of $B^\ast \bar{B}^\ast$, {while in the present work, we cannot find a bound state solution for this system due  to the consideration of the $\eta-\eta^{\prime}$ mixing.}

{A summary for the possible $\mathcal{P}^\ast-\mathcal{P}^\ast$ molecular state and the comparison with the estimation in the chiral and extended chiral quark model are presented in Table \ref{Tab:sum3}. The present calculations indicate there exist isoscalar bound states of the $D^\ast \bar{D}^\ast$ and $B^\ast \bar{B}^\ast$ with $J=0$, $1$ and $2$, while the estimations in Ref. \cite{Liu:2008mi} could only confirm the molecular state with $J=2$ for $D^\ast \bar{D}^\ast$ system and $J=1$ and $2$ for the $B^\ast \bar{B}^\ast$ system.}

\section{Summary \label{Sec:4}}

We have performed a systematic study of the possible molecular states composed of $S$ wave heavy-light mesons, where $S-D$ mixing and {$\eta-\eta^{\prime}$ mixing} are taken into consideration. From the present calculations and the comparison with the experimental observation, we can conclude:

1. Our calculation supports the $X(3872)$ as a loosely shallow $D\bar{D}^\ast + h.c $ molecular state with $I(J^{PC})=0(1^{++})$.

2. The counterpart of the $X(3872)$ in the bottom sector could be a molecular state composed of $B \bar{B}^\ast +h.c$.

3. The molecule assignments of the $Z_c(3900)$, $Z_c(4020)$, $Z_b(10610)$ and $Z_b(10650)$ are not supported by the present calculations.

4. We find three bound states composed of $D_{s}^{\ast + }D_s^{\ast -}$ with $J^{PC}=0^{++}, 1^{+-}$ and $2^{++}$, which is different from the quantum numbers of the $Y(4140)$ reported by the LHCb Collaboration. Thus, the $Y(4140)$ cannot be assigned as a molecular state composed of $D_s^{\ast +} D_s^{\ast -}$ in our calculations.

5. We predict more molecular states in the present calculations.
For the $\mathcal{P}-\mathcal{P}$ type, three molecular states,
$\Phi_8^0$, $\Omega_8^0$ and $\Omega_{s1}^0$, are predicted. In the
$\mathcal{P} -\mathcal{P}^\ast$ system, besides the $X(3872)$ and
its bottom counterpart, we also predict six new molecular states.

To summarize, in the present work, we have systematically {studied} the molecular states composed of the $S$ wave
heavy-light mesons, where the $S-D$ mixing  and {$\eta-\eta^{\prime}$ mixing} are explicitly
considered. In the present calculation, the observed $X(3872)$
could be interpreted as a loosely shallow $D\bar{D}^\ast +h.c$
molecular state, while $Z_c(3900)/ Z_c(4020)$ and
$Z_b(10610)$ and $Z_b(10650)$ cannot be molecular states. We
have also predicted some new molecular {states}, which could be searched for in
forthcoming experimental measurements.

\section*{Acknowledgements}
This project is supported by the National Natural Science Foundation
of China under Grant No. 11375240, No. 11565023.


\end{CJK*}

\begin{thebibliography}{99}


\bibitem{Chen:2016qju}
  H.~X.~Chen, W.~Chen, X.~Liu and S.~L.~Zhu,
  Phys.\ Rept.\  {\bf 639}, 1 (2016)

\bibitem{Choi:2003ue}
  S.~K.~Choi {\it et al.} [Belle Collaboration],
  Phys.\ Rev.\ Lett.\  {\bf 91}, 262001 (2003)

\bibitem{Abe:2005ix}
  K.~Abe {\it et al.}  [Belle Collaboration],
  arXiv:hep-ex/0505037.

\bibitem{Gokhroo:2006bt}
  G.~Gokhroo {\it et al.}  [Belle Collaboration],
  Phys.\ Rev.\ Lett.\  {\bf 97}, 162002 (2006)

\bibitem{:2008te}
  I.~Adachi {\it et al.}  [Belle Collaboration],
  arXiv:0809.1224 [hep-ex].

\bibitem{:2008su}
  I.~Adachi {\it et al.}  [Belle Collaboration],
  arXiv:0810.0358 [hep-ex].

\bibitem{Aubert:2004ns}
  B.~Aubert {\it et al.}  [BABAR Collaboration],
  Phys.\ Rev.\  D {\bf 71}, 071103 (2005)

\bibitem{Aubert:2004fc}
  B.~Aubert {\it et al.}  [BABAR Collaboration],
  Phys.\ Rev.\ Lett.\  {\bf 93}, 041801 (2004)

\bibitem{Aubert:2005eg}
  B.~Aubert {\it et al.}  [BABAR Collaboration],
  Phys.\ Rev.\  D {\bf 71}, 052001 (2005)

\bibitem{Aubert:2005zh}
  B.~Aubert {\it et al.}  [BABAR Collaboration],
  Phys.\ Rev.\  D {\bf 73}, 011101 (2006)

\bibitem{Aubert:2005vi}
  B.~Aubert {\it et al.}  [BABAR Collaboration],
  Phys.\ Rev.\ Lett.\  {\bf 96}, 052002 (2006)

\bibitem{Aubert:2006aj}
  B.~Aubert {\it et al.}  [BABAR Collaboration],
  Phys.\ Rev.\  D {\bf 74}, 071101 (2006)

\bibitem{Aubert:2007rva}
  B.~Aubert {\it et al.}  [BABAR Collaboration],
  Phys.\ Rev.\  D {\bf 77}, 011102 (2008)




\bibitem{Acosta:2003zx}
  D.~E.~Acosta {\it et al.}  [CDF II Collaboration],
  Phys.\ Rev.\ Lett.\  {\bf 93}, 072001 (2004)

\bibitem{Abulencia:2005zc}
  A.~Abulencia {\it et al.}  [CDF Collaboration],
  Phys.\ Rev.\ Lett.\  {\bf 96}, 102002 (2006)

\bibitem{Abulencia:2006ma}
  A.~Abulencia {\it et al.}  [CDF Collaboration],
  Phys.\ Rev.\ Lett.\  {\bf 98}, 132002 (2007)


\bibitem{Aaltonen:2009vj}
  T.~Aaltonen {\it et al.}  [CDF Collaboration],
  Phys.\ Rev.\ Lett.\  {\bf 103}, 152001 (2009)


\bibitem{Abazov:2004kp}
  V.~M.~Abazov {\it et al.}  [D0 Collaboration],
  Phys.\ Rev.\ Lett.\  {\bf 93}, 162002 (2004)


\bibitem{Aaij:2011sn}
  R.~Aaij {\it et al.} [LHCb Collaboration],
  Eur.\ Phys.\ J.\ C {\bf 72}, 1972 (2012)

\bibitem{Aaij:2013zoa}
  R.~Aaij {\it et al.} [LHCb Collaboration],
  Phys.\ Rev.\ Lett.\  {\bf 110}, 222001 (2013)


\bibitem{Aaij:2014ala}
  R.~Aaij {\it et al.} [LHCb Collaboration],
  Nucl.\ Phys.\ B {\bf 886}, 665 (2014)

\bibitem{Aaij:2015eva}
  R.~Aaij {\it et al.} [LHCb Collaboration],
  Phys.\ Rev.\ D {\bf 92}, no. 1, 011102 (2015)

\bibitem{Ablikim:2013dyn}
  M.~Ablikim {\it et al.} [BESIII Collaboration],
  Phys.\ Rev.\ Lett.\  {\bf 112} (2014) no.9,  092001

\bibitem{Aubert:2004zr}
  B.~Aubert {\it et al.}  [BaBar Collaboration],
  Phys.\ Rev.\  D {\bf 71}, 031501 (2005)

\bibitem{Ablikim:2013mio}
  M.~Ablikim {\it et al.} [BESIII Collaboration],
  Phys.\ Rev.\ Lett.\  {\bf 110}, 252001 (2013)

\bibitem{Liu:2013dau}
  Z.~Q.~Liu {\it et al.} [Belle Collaboration],
  Phys.\ Rev.\ Lett.\  {\bf 110}, 252002 (2013)


\bibitem{Xiao:2013iha}
  T.~Xiao, S.~Dobbs, A.~Tomaradze and K.~K.~Seth,
  Phys.\ Lett.\ B {\bf 727}, 366 (2013)

\bibitem{Ablikim:2013xfr}
  M.~Ablikim {\it et al.} [BESIII Collaboration],
  Phys.\ Rev.\ Lett.\  {\bf 112}, no. 2, 022001 (2014)


\bibitem{Ablikim:2015tbp}
  M.~Ablikim {\it et al.} [BESIII Collaboration],
  Phys.\ Rev.\ Lett.\  {\bf 115}, 112003 (2015).


\bibitem{Ablikim:2015gda}
  M.~Ablikim {\it et al.},
  arXiv:1509.05620 [hep-ex].


\bibitem{Ablikim:2013wzq}
  M.~Ablikim {\it et al.} [BESIII Collaboration],
  Phys.\ Rev.\ Lett.\  {\bf 111}, no. 24, 242001 (2013)

\bibitem{Ablikim:2014dxl}
  M.~Ablikim {\it et al.} [BESIII Collaboration],
  Phys.\ Rev.\ Lett.\  {\bf 113} (2014) no.21,  212002

\bibitem{Ablikim:2013emm}
  M.~Ablikim {\it et al.} [BESIII Collaboration],
  Phys.\ Rev.\ Lett.\  {\bf 112} (2014) no.13,  132001

\bibitem{Ablikim:2015vvn}
  M.~Ablikim {\it et al.} [BESIII Collaboration],
  Phys.\ Rev.\ Lett.\  {\bf 115} (2015) no.18,  182002


\bibitem{Aaltonen:2009tz}
  T.~Aaltonen {\it et al.} [CDF Collaboration],
  Phys.\ Rev.\ Lett.\  {\bf 102}, 242002 (2009).


\bibitem{Aaltonen:2011at}
  T.~Aaltonen {\it et al.} [CDF Collaboration],
  arXiv:1101.6058 [hep-ex].



\bibitem{Chatrchyan:2013dma}
  S.~Chatrchyan {\it et al.} [CMS Collaboration],
  Phys.\ Lett.\ B {\bf 734}, 261 (2014).

\bibitem{Abazov:2013xda}
  V.~M.~Abazov {\it et al.} [D0 Collaboration],
  Phys.\ Rev.\ D {\bf 89}, 012004 (2014).



\bibitem{Aaij:2016iza}
  R.~Aaij {\it et al.} [LHCb Collaboration],
  arXiv:1606.07895 [hep-ex].


\bibitem{Aaij:2016nsc}
  R.~Aaij {\it et al.} [LHCb Collaboration],
  arXiv:1606.07898 [hep-ex].



\bibitem{Belle:2011aa}
  A.~Bondar {\it et al.} [Belle Collaboration],
  Phys.\ Rev.\ Lett.\  {\bf 108} (2012) 122001


\bibitem{Garmash:2014dhx}
  A.~Garmash {\it et al.} [Belle Collaboration],
  Phys.\ Rev.\ D {\bf 91} (2015) no.7,  072003


\bibitem{Adachi:2012cx}
  I.~Adachi {\it et al.} [Belle Collaboration],
  arXiv:1209.6450 [hep-ex].


\bibitem{Garmash:2015rfd}
  A.~Garmash {\it et al.} [Belle Collaboration],
  Phys.\ Rev.\ Lett.\  {\bf 116} (2016) no.21,  212001


\bibitem{Krokovny:2013mgx}
  P.~Krokovny {\it et al.} [Belle Collaboration],
  Phys.\ Rev.\ D {\bf 88} (2013) no.5,  052016

\bibitem{Abdesselam:2015zza}
  A.~Abdesselam {\it et al.} [Belle Collaboration],
  Phys.\ Rev.\ Lett.\  {\bf 117} (2016) no.14,  142001

\bibitem{Hou:2006it}
  W.~S.~Hou,
  Phys.\ Rev.\ D {\bf 74}, 017504 (2006)


\bibitem{Li:2015uwa}
  G.~Li and Z.~Zhou,
  Phys.\ Rev.\ D {\bf 91} (2015) no.3,  034020

\bibitem{Wu:2016dws}
  Q.~Wu, G.~Li, F.~Shao, Q.~Wang, R.~Wang, Y.~Zhang and Y.~Zheng,
  Adv.\ High Energy Phys.\  {\bf 2016}, 3729050 (2016)

\bibitem{He:2014sqj}
  X.~H.~He {\it et al.} [Belle Collaboration],
  Phys.\ Rev.\ Lett.\  {\bf 113} (2014) no.14,  142001


\bibitem{Kalashnikova:2005ui}
  Y.~S.~Kalashnikova,
  Phys.\ Rev.\ D {\bf 72}, 034010 (2005)

\bibitem{Zhang:2009bv}
  O.~Zhang, C.~Meng and H.~Q.~Zheng,
  Phys.\ Lett.\ B {\bf 680}, 453 (2009)

\bibitem{Kalashnikova:2009gt}
  Y.~S.~Kalashnikova and A.~V.~Nefediev,
  Phys.\ Rev.\ D {\bf 80}, 074004 (2009)

\bibitem{Li:2009zu}
  B.~Q.~Li and K.~T.~Chao,
  Phys.\ Rev.\ D {\bf 79}, 094004 (2009)

\bibitem{Ferretti:2013faa}
  J.~Ferretti, G.~Galat¨¤ and E.~Santopinto,
  Phys.\ Rev.\ C {\bf 88}, 015207 (2013).

\bibitem{Eichten:2004uh}
  E.~J.~Eichten, K.~Lane and C.~Quigg,
  Phys.\ Rev.\ D {\bf 69}, 094019 (2004).


\bibitem{Pennington:2007xr}
  M.~R.~Pennington and D.~J.~Wilson,
  Phys.\ Rev.\ D {\bf 76}, 077502 (2007).



\bibitem{Chen:2016iua}
  D.~Y.~Chen,
  arXiv:1611.00109 [hep-ph].

\bibitem{Maiani:2004vq}
  L.~Maiani, F.~Piccinini, A.~D.~Polosa and V.~Riquer,
  Phys.\ Rev.\ D {\bf 71} (2005) 014028


\bibitem{Chiu:2006us}
  T.~W.~Chiu {\it et al.} [TWQCD Collaboration],
  Phys.\ Rev.\ D {\bf 73} (2006) 111503
   Erratum: [Phys.\ Rev.\ D {\bf 75} (2007) 019902]


\bibitem{Ebert:2005nc}
  D.~Ebert, R.~N.~Faustov and V.~O.~Galkin,
  Phys.\ Lett.\ B {\bf 634} (2006) 214



\bibitem{Ali:2014dva}
  A.~Ali, L.~Maiani, A.~D.~Polosa and V.~Riquer,
  Phys.\ Rev.\ D {\bf 91}, no. 1, 017502 (2015)


\bibitem{Ali:2013xba}
  A.~Ali, C.~Hambrock and W.~Wang,
  Phys.\ Rev.\ D {\bf 88}, no. 5, 054026 (2013)


\bibitem{Chen:2010ze}
  W.~Chen and S.~L.~Zhu,
  Phys.\ Rev.\ D {\bf 83}, 034010 (2011)

\bibitem{Chen:2015ata}
  W.~Chen, T.~G.~Steele, H.~X.~Chen and S.~L.~Zhu,
  Phys.\ Rev.\ D {\bf 92}, no. 5, 054002 (2015)


\bibitem{Chen:2013omd}
  W.~Chen, T.~G.~Steele, M.~L.~Du and S.~L.~Zhu,
  Eur.\ Phys.\ J.\ C {\bf 74}, no. 2, 2773 (2014)


\bibitem{Wang:2013llv}
  Z.~G.~Wang,
  Commun.\ Theor.\ Phys.\  {\bf 63}, no. 4, 466 (2015)


\bibitem{Qiao:2013dda}
  C.~F.~Qiao and L.~Tang,
  Eur.\ Phys.\ J.\ C {\bf 74}, 2810 (2014)


\bibitem{Wang:2015nwa}
  Z.~G.~Wang,
  Int.\ J.\ Mod.\ Phys.\ A {\bf 30}, no. 30, 1550168 (2015)


\bibitem{Bugg:2011jr}
  D.~V.~Bugg,
  Europhys.\ Lett.\  {\bf 96}, 11002 (2011)



\bibitem{Bugg:2011ub}
  D.~V.~Bugg,
  arXiv:1101.1659 [hep-ph].

\bibitem{Chen:2011pv}
  D.~Y.~Chen and X.~Liu,
  Phys.\ Rev.\ D {\bf 84} (2011) 094003
  [arXiv:1106.3798 [hep-ph]].

\bibitem{Chen:2013coa}
  D.~Y.~Chen, X.~Liu and T.~Matsuki,
  Phys.\ Rev.\ D {\bf 88} (2013) no.3,  036008

\bibitem{Chen:2011xk}
  D.~Y.~Chen and X.~Liu,
  Phys.\ Rev.\ D {\bf 84}, 034032 (2011)

\bibitem{Chen:2011zv}
  D.~Y.~Chen, X.~Liu and S.~L.~Zhu,
  Phys.\ Rev.\ D {\bf 84} (2011) 074016

\bibitem{Chen:2013wca}
  D.~Y.~Chen, X.~Liu and T.~Matsuki,
  Phys.\ Rev.\ Lett.\  {\bf 110} (2013) no.23,  232001



\bibitem{Swanson:2003tb}
  E.~S.~Swanson,
  Phys.\ Lett.\ B {\bf 588}, 189 (2004)









\bibitem{Liu:2008fh}
  Y.~R.~Liu, X.~Liu, W.~Z.~Deng and S.~L.~Zhu,
  Eur.\ Phys.\ J.\ C {\bf 56}, 63 (2008)


\bibitem{Thomas:2008ja}
  C.~E.~Thomas and F.~E.~Close,
  Phys.\ Rev.\ D {\bf 78}, 034007 (2008)


\bibitem{Liu:2008tn}X.~Liu, Z.~G.~Luo, Y.~R.~Liu and S.~L.~Zhu,
  Eur.\ Phys.\ J.\ C {\bf 61}, 411 (2009)


\bibitem{AlFiky:2005jd}
  M.~T.~AlFiky, F.~Gabbiani and A.~A.~Petrov,
  Phys.\ Lett.\ B {\bf 640}, 238 (2006)


\bibitem{Lee:2009hy}
  I.~W.~Lee, A.~Faessler, T.~Gutsche and V.~E.~Lyubovitskij,
  Phys.\ Rev.\ D {\bf 80}, 094005 (2009)


\bibitem{Chen:2013pya}
  W.~Chen, H.~y.~Jin, R.~T.~Kleiv, T.~G.~Steele, M.~Wang and Q.~Xu,
  Phys.\ Rev.\ D {\bf 88}, no. 4, 045027 (2013)








\bibitem{Dong:2009yp}
  Y.~Dong, A.~Faessler, T.~Gutsche, S.~Kovalenko and V.~E.~Lyubovitskij,
  Phys.\ Rev.\ D {\bf 79}, 094013 (2009)

\bibitem{Dong:2014zka}
  Y.~Dong, A.~Faessler, T.~Gutsche and V.~E.~Lyubovitskij,
  Phys.\ Rev.\ D {\bf 90}, no. 7, 074032 (2014)


\bibitem{Harada:2010bs}
  M.~Harada and Y.~L.~Ma,
  Prog.\ Theor.\ Phys.\  {\bf 126}, 91 (2011)

\bibitem{Braaten:2010mg}
  E.~Braaten, H.-W.~Hammer and T.~Mehen,
  Phys.\ Rev.\ D {\bf 82}, 034018 (2010)






























\bibitem{Aceti:2014uea}
  F.~Aceti, M.~Bayar, E.~Oset, A.~Martinez Torres, K.~P.~Khemchandani, J.~M.~Dias, F.~S.~Navarra and M.~Nielsen,
  Phys.\ Rev.\ D {\bf 90}, no. 1, 016003 (2014)


\bibitem{Sun:2011uh}
  Z.~F.~Sun, J.~He, X.~Liu, Z.~G.~Luo and S.~L.~Zhu,
  Phys.\ Rev.\ D {\bf 84}, 054002 (2011)

\bibitem{Sun:2012zzd}
  Z.~F.~Sun, Z.~G.~Luo, J.~He, X.~Liu and S.~L.~Zhu,
  Chin.\ Phys.\ C {\bf 36}, 194 (2012).


\bibitem{Wang:2013daa}
  Z.~G.~Wang and T.~Huang,
  Eur.\ Phys.\ J.\ C {\bf 74}, no. 5, 2891 (2014)

\bibitem{Wu:2016ypc}
  Q.~Wu, G.~Li, F.~Shao and R.~Wang,
  Phys.\ Rev.\ D {\bf 94}, no. 1, 014015 (2016).

\bibitem{Li:2013xia}
  G.~Li,
  Eur.\ Phys.\ J.\ C {\bf 73}, no. 11, 2621 (2013)






\bibitem{Dong:2013iqa}
  Y.~Dong, A.~Faessler, T.~Gutsche and V.~E.~Lyubovitskij,


\bibitem{Dong:2013kta}
  Y.~Dong, A.~Faessler, T.~Gutsche and V.~E.~Lyubovitskij,
  Phys.\ Rev.\ D {\bf 89}, 034018 (2014).


\bibitem{Li:2014pfa}
  G.~Li, X.~H.~Liu and Z.~Zhou,
  Phys.\ Rev.\ D {\bf 90}, 054006 (2014).

\bibitem{Gutsche:2014zda}
  T.~Gutsche, M.~Kesenheimer and V.~E.~Lyubovitskij,
  Phys.\ Rev.\ D {\bf 90},  094013 (2014).

\bibitem{Esposito:2014hsa}
  A.~Esposito, A.~L.~Guerrieri and A.~Pilloni,
  Phys.\ Lett.\ B {\bf 746}, 194 (2015).

\bibitem{Ke:2013gia}
  H.~W.~Ke, Z.~T.~Wei and X.~Q.~Li,
  Eur.\ Phys.\ J.\ C {\bf 73}, 2561 (2013).


\bibitem{Chen:2015igx}
  D.~Y.~Chen and Y.~B.~Dong,
  Phys.\ Rev.\ D {\bf 93}, no. 1, 014003 (2016)




\bibitem{Mahajan:2009pj}
  N.~Mahajan,
  Phys.\ Lett.\ B {\bf 679}, 228 (2009).


\bibitem{Ding:2009vd}
  G.~J.~Ding,
  Eur.\ Phys.\ J.\ C {\bf 64}, 297 (2009).



\bibitem{Molina:2009ct}
  R.~Molina and E.~Oset,
  Phys.\ Rev.\ D {\bf 80}, 114013 (2009).

\bibitem{Liu:2009pu}
  X.~Liu and H.~W.~Ke,
  Phys.\ Rev.\ D {\bf 80}, 034009 (2009).

\bibitem{Zhang:2009vs}
  J.~R.~Zhang and M.~Q.~Huang,
  Phys.\ Rev.\ D {\bf 80}, 056004 (2009).



\bibitem{Liu:2009ei}
  X.~Liu and S.~L.~Zhu,
  Phys.\ Rev.\ D {\bf 80}, 017502 (2009)
  Erratum: [Phys.\ Rev.\ D {\bf 85}, 019902 (2012)]

\bibitem{Albuquerque:2009ak}
  R.~M.~Albuquerque, M.~E.~Bracco and M.~Nielsen,
  Phys.\ Lett.\ B {\bf 678}, 186 (2009).

\bibitem{Zhang:2009st}
  J.~R.~Zhang and M.~Q.~Huang,
  J.\ Phys.\ G {\bf 37}, 025005 (2010).




\bibitem{Wang:2014gwa}
  Z.~G.~Wang,
  Eur.\ Phys.\ J.\ C {\bf 74}, 2963 (2014).



\bibitem{Chen:2016ugz}
  X.~Chen, X.~L¨¹, R.~Shi and X.~Guo,
  Nucl.\ Phys.\ B {\bf 909}, 243 (2016).









\bibitem{Branz:2009yt}
  T.~Branz, T.~Gutsche and V.~E.~Lyubovitskij,
  Phys.\ Rev.\ D {\bf 80}, 054019 (2009)

\bibitem{Li:2012wf}
  M.~T.~Li, W.~L.~Wang, Y.~B.~Dong and Z.~Y.~Zhang,
  J.\ Phys.\ G {\bf 40}, 015003 (2013)

\bibitem{Yang:2011rp}
  Y.~Yang, J.~Ping, C.~Deng and H.~S.~Zong,
  J.\ Phys.\ G {\bf 39}, 105001 (2012)


\bibitem{Zhang:2011jja}
  J.~R.~Zhang, M.~Zhong and M.~Q.~Huang,
  Phys.\ Lett.\ B {\bf 704}, 312 (2011)


\bibitem{Dong:2012hc}
  Y.~Dong, A.~Faessler, T.~Gutsche and V.~E.~Lyubovitskij,
  J.\ Phys.\ G {\bf 40}, 015002 (2013)




\bibitem{Li:2012as}
  G.~Li, F.~l.~Shao, C.~W.~Zhao and Q.~Zhao,
  Phys.\ Rev.\ D {\bf 87}, no. 3, 034020 (2013)


\bibitem{Cleven:2013sq}
  M.~Cleven, Q.~Wang, F.~K.~Guo, C.~Hanhart, U.~G.~Meissner and Q.~Zhao,
  Phys.\ Rev.\ D {\bf 87}, no. 7, 074006 (2013)

\bibitem{Li:2012uc}
  X.~Li and M.~B.~Voloshin,
  Phys.\ Rev.\ D {\bf 86}, 077502 (2012)


\bibitem{Mehen:2011yh}
  T.~Mehen and J.~W.~Powell,
  Phys.\ Rev.\ D {\bf 84}, 114013 (2011)


\bibitem{Carlson:1997qn}
  J.~Carlson and R.~Schiavilla,
  Rev.\ Mod.\ Phys.\  {\bf 70}, 743 (1998).

\bibitem{Arnold:1980zj}
  R.~G.~Arnold, C.~E.~Carlson and F.~Gross,
  Phys.\ Rev.\ C {\bf 23} (1981) 363.



















\bibitem{Casalbuoni:1996pg}R.~Casalbuoni, A.~Deandrea, N.~Di Bartolomeo, R.~Gatto, F.~Feruglio and G.~Nardulli,
  Phys.\ Rept.\  {\bf 281}, 145 (1997)


\bibitem{Cheng:1992xi} H.~Y.~Cheng, C.~Y.~Cheung, G.~L.~Lin, Y.~C.~Lin, T.~M.~Yan and H.~L.~Yu,
  Phys.\ Rev.\ D {\bf 47}, 1030 (1993)

\bibitem{Yan:1992gz}T.~M.~Yan, H.~Y.~Cheng, C.~Y.~Cheung, G.~L.~Lin, Y.~C.~Lin and H.~L.~Yu,
  Phys.\ Rev.\ D {\bf 46}, 1148 (1992)

\bibitem{Wise:1992hn}
M.~B.~Wise,
  Phys.\ Rev.\ D {\bf 45}, 2188 (1992).

\bibitem{Burdman:1992gh} G.~Burdman and J.~F.~Donoghue,
  Phys.\ Lett.\ B {\bf 280}, 287 (1992).

\bibitem{Falk:1992cx}A.~F.~Falk and M.~E.~Luke,
  Phys.\ Lett.\ B {\bf 292}, 119 (1992)


  \bibitem{Isola:2003fh} C.~Isola, M.~Ladisa, G.~Nardulli and P.~Santorelli,
  Phys.\ Rev.\ D {\bf 68}, 114001 (2003)


\bibitem{Bando:1987br}
  M.~Bando, T.~Kugo and K.~Yamawaki,
  Phys.\ Rept.\  {\bf 164}, 217 (1988).



\bibitem{Coffman:1988ve}
  D.~Coffman {\it et al.}  [MARK-III Collaboration],
  Phys.\ Rev.\ D {\bf 38} (1988) 2695
   [Erratum-ibid.\ D {\bf 40} (1989) 3788].


\bibitem{Jousset:1988ni}
  J.~Jousset {\it et al.}  [DM2 Collaboration],
  Phys.\ Rev.\ D {\bf 41} (1990) 1389.


\bibitem{Tornqvist:1993vu}N.~A.~Tornqvist,
  Nuovo Cim.\ A {\bf 107}, 2471 (1994)

\bibitem{Tornqvist:1993ng} N.~A.~Tornqvist,
  Z.\ Phys.\ C {\bf 61}, 525 (1994)

\bibitem{Locher:1993cc}
M.~P.~Locher, Y.~Lu and B.~S.~Zou,
  Z.\ Phys.\ A {\bf 347}, 281 (1994)

\bibitem{Li:1996yn}
 X.~Q.~Li, D.~V.~Bugg and B.~S.~Zou,
  Phys.\ Rev.\ D {\bf 55}, 1421 (1997).










\bibitem{Moiseyev:1988nm}
N. Moiseyev Phys.\ Rept.\  {\bf 302}, 212-293 (1988).

\bibitem{Guo:2010cpc}
J.-Y. Guo, M. Yu, J. Wang, B.-M. Yao, P. Jiao, Comput. Phys.
Commun {\bf181}, 550 (2010).

\bibitem{Liu:2013lq}
Q. Liu, Z.-M. Niu, and J.-Y. Guo, Phys.\ Rev.\ A {\bf87}, 052122
(2013) (2013).

\bibitem{Guo:2010zzm}
J.~Y.~Guo, X.~Z.~Fang, P.~Jiao, J.~Wang and B.~M.~Yao,
  Phys.\ Rev.\ C {\bf 82}, 034318 (2010).

\bibitem{Liu:2012da}
Q.~Liu, J.~Y.~Guo, Z.~M.~Niu and S.~W.~Chen,
  Phys.\ Rev.\ C {\bf 86}, 054312 (2012)

\bibitem{Alhaidari:2001AD}
A. D. Alhaidari, H. A. Yamani and M. S. Abdelmonem, Phys.\ Rev.\ A
{\bf63},062708 (2001)


\bibitem{Liu:2008mi}
  Y.~R.~Liu and Z.~Y.~Zhang,
  Phys.\ Rev.\ C {\bf 80}, 015208 (2009)

\bibitem{Li:2012cs}
  N.~Li and S.~L.~Zhu,
  Phys.\ Rev.\ D {\bf 86}, 074022 (2012)


\bibitem{He:2015mja}
  J.~He,
  Phys.\ Rev.\ D {\bf 92}, no. 3, 034004 (2015)



\bibitem{Ohkoda:2011vj}
  S.~Ohkoda, Y.~Yamaguchi, S.~Yasui, K.~Sudoh and A.~Hosaka,
  Phys.\ Rev.\ D {\bf 86}, 014004 (2012)


\end{thebibliography}
\end{document}